\documentclass[sn-mathphys]{sn-jnl}% Math and Physical Sciences Reference Style
\jyear{2021}%

\theoremstyle{thmstyleone}%
\newtheorem{theorem}{Theorem}%  meant for continuous numbers

\theoremstyle{thmstyletwo}%

\theoremstyle{thmstylethree}%

\begin{document}

\title[Factor Augmented Quantile Regression Model]{Factor Augmented Quantile Regression Model}


\author[1]{\fnm{Xiaoyang} \sur{Wei}}

\author*[1]{\fnm{Yanlin} \sur{Tang}}\email{yltang@fem.ecnu.edu.cn}

\author[2]{\fnm{Xu} \sur{Guo}}

\author[3]{\fnm{Meiling} \sur{Hao}}

\author[2]{\fnm{Yanmei} \sur{Shi}}
\affil[1]{KLATASDS-MOE, School of Statistics, East China Normal University, Shanghai, China}

\affil[2]{School of Statistics, Beijing Normal University, Beijing, China}

\affil[3]{School of Statistics, University of International Business and Economics, Beijing, China}


\abstract{
Along with the widespread adoption of high-dimensional data, traditional statistical methods face significant challenges in handling problems with high correlation of variables, heavy-tailed distribution, and coexistence of sparse and dense effects. In this paper, we propose a factor-augmented quantile regression (FAQR) framework to address these challenges simultaneously within a unified framework. The proposed FAQR combines the robustness of quantile regression and the ability of factor analysis to effectively capture dependencies among high-dimensional covariates, and also provides a framework to capture dense effects (through common factors) and sparse effects (through idiosyncratic components) of the covariates. 
To overcome the lack of smoothness of the quantile loss function, convolution smoothing is introduced, which not only improves computational efficiency but also eases theoretical derivation. Theoretical analysis establishes the accuracy of factor selection and consistency in parameter estimation under mild regularity conditions. Furthermore, we develop a Bootstrap-based diagnostic procedure to assess the adequacy of the factor model. Simulation experiments verify the rationality of FAQR in different noise scenarios such as normal and $t_2$ distributions.}



\maketitle

\section{Introduction}

Rapid advancement of data collection technologies has led to a surge of high-dimensional datasets across diverse domains, including genomic studies, image processing, and text mining. These high-dimensional datasets frequently present three fundamental challenges: (1) inherent multicollinearity among variables that undermines classical statistical methods, (2) heavy-tailed distributions violating Gaussian assumptions, and (3) heterogeneous relationships between covariates and responses ranging from sparse (where only a few covariates are relevant) to dense (where many features contribute jointly), requiring methods that flexibly adapt to both regimes. Traditional methods such as ordinary least squares (OLS) prove inadequate in addressing these intertwined challenges, requiring more sophisticated analytical frameworks. This phenomenon is exemplified, for example, in motivated macroeconomic research utilizing the Federal Reserve's comprehensive FRED-MD database, in Section \ref{real}. Time-series data within FRED-MD often exhibit heavy-tailed distributions, and the extensive coverage of hundreds of economic indicators inevitably leads to significant multicollinearity among variables. The question of how to flexibly and comprehensively deal with these problems in a single model is an important challenge in data analysis, while existing approaches address only one or two of these challenges.

In the existing literature, there are two types of methods to model the relationship between response and high-dimensional covariates. The first approach is sparse linear model, where penalized regression is usually used to select important covariates, say the lasso method \citep{tibshirani1996regression}, adaptive lasso \citep{zou2006adaptive}, and SCAD \citep{fan2001variable}. The second approach is factor analysis, which solves the challenges of highly collinear and high-dimensional data by representing covariates through a lower-dimensional set of latent factors \citep{carhart1997persistence, bai2006confidence, fan2017sufficient}. Its theoretical work has also been extensively studied \citep{bai2003inferential, fan2013large, ouyang2023high}. However, existing approaches rarely unify sparse and dense effects within one framework. One related work is Fan et al. \cite{fan2020factor}, where factor-adjusted regularization is proposed to improve the performance of model selection in a highly dimensional sparse linear model.

Recently, to simultaneously capture the sparse and dense effects of the covariates, Fan et al. \cite{fan2023latent} 
developed the Factor Augmented sparse linear Regression Model (FARM), which incorporates both common factors and idiosyncratic components in the modeling. Traditional factor models typically include only $\boldsymbol{f}$, the common factors extracted, as covariates after decomposing the feature vector $\boldsymbol{X}$. 
Although $\boldsymbol{f}$ may contribute significantly to the response, it is unlikely that they possess complete explanatory power, and the idiosyncratic components $\boldsymbol{u}$ may retain important predictive information, especially with respect to individual-specific effects. Compared to the conventional high-dimensional sparse linear regression model, the Factor Augmented sparse linear Regression Model (FARM) utilizes common factors more effectively, especially when the feature vector $\boldsymbol{X}$ exhibits high collinearity and sparse regression tends to become unstable. To demonstrate its necessity, they systematically compared FARM with the traditional factor regression model and the high-dimensional sparse regression model.
Therefore, FARM is sufficiently versatile to connect dimensionality reduction with sparse regression, enabling the unified handling of sparse and dense effects within a single framework. 
However, FARM performs poorly when the response variable is subject to heavy tails.

Unlike the ordinary least squares (OLS) framework adopted in FARM, quantile regression (QR, \cite{koenker1978regression, koenker2018handbook}) is robust to outliers and provides a more comprehensive understanding of the entire conditional distribution of response variables. In this paper, we propose a novel Factor Augmented Quantile Regression (FAQR) framework (\ref{e5}), which enables robust estimation of conditional quantiles while mitigating multicollinearity via factor analysis and thus integrates sparse and dense predictor-response relationships into a unified model.
The novelty of FAQR lies in three key aspects. First, it performs joint estimation and dimension reduction within a single framework, resolving complexity in high-dimensional covariate systems. Second, it employs convolution smoothing to address the non-differentiability of traditional quantile regression loss function, preserving quantile-specific properties with a differentiable alternative. Although the optimization problem in QR can be transformed into a linear program, which requires substantial computing resources when applied to high-dimensional datasets. In this study, we adopt the convolution-smoothed QR to achieve improved computational efficiency without compromising statistical validity. Third, for optimization, we adopt a modified local adaptive majorize-minimization (LAMM) algorithm with step size adaptation \citep{tan2022high}, which achieves efficient computation and optimal convergence under standard regularity conditions, avoiding the computational burden of linear programming in high dimensions.

The primary contributions of this work are summarized as follows:
\begin{enumerate}
\item \textbf{Unified framework.} We develop a high-dimensional quantile regression model that integrates latent factor analysis with sparse regularization, jointly addressing multicollinearity and heavy-tailed distributions in a single step. The FAQR provides a complete picture of the relationship between the response variable and feature vector, allowing heterogeneity in the data.
\item \textbf{Theoretical guarantees.} We establish $\ell_2$ consistency and convergence rates for estimates of both common factor and idiosyncratic parameters under standard regularity conditions.
\item \textbf{Statistical inference.} Through extensive numerical simulations, we demonstrate the finite-sample performance of our hypothesis testing method, including accuracy and power under varying signal strengths and error distributions.
\end{enumerate}

The rest of the paper is organized as follows. We propose the FAQR model and describe its estimation procedure in Section 2, present the consistency of the estimator in Section 3, and provide a bootstrap method to test the significance of the idiosyncratic components in Section 4. In Section 5, we conduct simulation studies for the accuracy and adequacy test of the factor model. In Section 6, we apply the proposed method to the motivated example. A discussion is provided in Section 7. All technique proofs are deferred to the online Supplementary Materials.

\textbf{Notation}. Let $\mathbb{I}(\cdot)$ denote the indicator function. For a vector $\boldsymbol{a}=\left(a_1, \ldots, a_m\right)^{\top} \in \mathbb{R}^m$, we denote its $\ell_q$ norm as $\|\boldsymbol{a}\|_q=\left(\sum_{\ell=1}^m\mid a_{\ell}\mid ^q\right)^{1 / q}, 1 \leq q<\infty,\|\boldsymbol{a}\|_{\infty}=\max _{1 \leq \ell \leq m}\mid a_{\ell}\mid $, and $\|\boldsymbol{a}\|_0=\sum_{\ell=1}^m \mathbb{I}\left(a_{\ell} \neq 0\right)$. For any integer $m$, we define $[m]=\{1, \ldots, m\}$. The Orlicz norm of a scalar random variable $X$ is defined as $\|X\|_{\psi_2}=\inf \left\{c>0: \mathbb{E} \exp \left(X^2 / c^2\right) \leq 2\right\}$. We define $\|\boldsymbol{A}\|_{\mathbb{F}}=\sqrt{\sum_{j k} A_{j k}^2},\|\boldsymbol{A}\|_{\max }=$ $\max _{j, k}\mid A_{j k}\mid ,\|\boldsymbol{A}\|_{\infty}=\max _j \sum_k\mid A_{j k}\mid$ and $\|\boldsymbol{A}\|_1=\max _k \sum_j\mid A_{j k}\mid$ to be its Frobenius norm, element-wise max-norm, matrix $\ell_{\infty}$-norm and matrix $\ell_1$-norm, respectively. Furthermore, if $a_n=O\left(b_n\right)$ is satisfied, we write $a_n \lesssim b_n$. If $a_n \lesssim b_n$ and $b_n \lesssim a_n$, we write it as $a_n \asymp b_n$ for short. In addition, let $a_n=O_{\mathbb{P}}\left(b_n\right)$ denote $\operatorname{Pr}\left(\mid a_n / b_n\mid \leq c\right) \rightarrow 1$ for some constant $c<\infty$. Let $a_n=o_{\mathbb{P}}\left(b_n\right)$ denote $\operatorname{Pr}\left(\mid a_n / b_n\mid >c\right) \rightarrow 0$ for any constant $c>0$. The parameters $c, c_0, C, C_1, C_2$ and $K^{\prime}$ appearing in this paper are all positive constants. In the following, we use $d_n$ and $s_n$ to represent the dimension and number of nonzero components of $\boldsymbol{\beta}^{*}(\tau)$, respectively. We use $\lambda_n$ for the tuning parameter of the $\ell_1$-penalty and $h_n$ for the smoothing bandwidth. For notational simplicity, we often omit the subscript $n$ in $d_n$, $s_n$, $\lambda_n$, and $h_n$ where the dependence on the sample size is clear from the context.

% Specifically, we assume that the observed $d$-dimensional covariate vector $\boldsymbol{x}$ conforms to the model:
% \begin{align}\label{eq2}
%     \boldsymbol{x}=\boldsymbol{B} \boldsymbol{f}+\boldsymbol{u},
% \end{align}
% where $\boldsymbol{f}$ denotes a $M$-dimensional vector of latent factors, $\boldsymbol{B} \in \mathbb{R}^{d \times M}$ represents the corresponding factor loading matrix, and $\boldsymbol{u}$ is a $d$-dimensional vector representing the idiosyncratic component, which is assumed to be uncorrelated with $\boldsymbol{f}$.

% To address the limitations of conventional factor models that assume purely dense effects, we extend the FARM framework \citep{fan2023latent} to quantile regression settings. 

\section{Factor Augmented Convolution Smoothing Quantile
Linear Model}

\subsection{Factor Augmented Quantile Regression Model}

A foundational framework for analyzing high-dimensional data with correlated features is the \textbf{F}actor \textbf{A}ugmented sparse linear \textbf{R}egression \textbf{M}odel (FARM) proposed by Fan et al. \cite{fan2023latent}. 
FARM integrates dimension reduction and sparse regression by incorporating both common factors $\boldsymbol{f}$ and idiosyncratic components $\boldsymbol{u}$ into a unified linear model for the response $Y$:
\begin{align*}
Y &= \boldsymbol{f}^{\top}\boldsymbol{\gamma}^{*} + \boldsymbol{u}^{\top}\boldsymbol{\beta}^{*} + \varepsilon, \\
\boldsymbol{X} &= \boldsymbol{B} \boldsymbol{f} + \boldsymbol{u}.
\end{align*}
Equivalently, FARM can be expressed as,
$$
\begin{aligned}
& Y=\boldsymbol{f}^{\top} \boldsymbol{\varphi}^{*}+\boldsymbol{x}^{\top} \boldsymbol{\beta}^{*}+\varepsilon, \\
& \boldsymbol{x}=\boldsymbol{B} \boldsymbol{f}+\boldsymbol{u},
\end{aligned}
$$
where $\boldsymbol{\varphi}^{*}=\boldsymbol{\gamma}^{*}-\boldsymbol{B}^{\top} \boldsymbol{\beta}^{*} \in \mathbb{R}^M$ quantifies the extra contribution of the M-dimensional latent factors $\boldsymbol{f}$ beyond the observed predictor $\boldsymbol{X}$. Here, $\boldsymbol{u}$ represents idiosyncratic component vector and $\boldsymbol{B}$ is the factor loading matrix. FARM notably encompasses both latent factor regression ($\boldsymbol{\beta}^{*}=0$) and sparse linear regression ($\boldsymbol{\varphi}^{*}=0$) as special cases. 

However, FARM is based on mean regression, thus it is sensitive to heavy-tailed errors. For robust estimation, we extend the FARM framework to quantile regression, developing the \textbf{F}actor \textbf{A}ugmented sparse linear \textbf{Q}uantile \textbf{R}egression model (FAQR). Suppose that we have $n$ independent and identically distributed (i.i.d.) random samples $\left\{\left(\boldsymbol{X}_{i}, Y_{i}\right)\right\}_{i=1}^n$ from the joint distribution of $(\boldsymbol{X}, Y)$, adhering to the following relationships:
\begin{align} \label{e5}
    \boldsymbol{X}_i&=\boldsymbol{B} \boldsymbol{f}_i+\boldsymbol{u}_i, \notag\\ 
Q_\tau\left(Y_{i} \mid \boldsymbol{f}_i, \boldsymbol{u}_i\right)&=\boldsymbol{f}_i^{\top} \boldsymbol{\gamma}^{*}(\tau)+\boldsymbol{u}_i^{\top} \boldsymbol{\beta}^{*}(\tau), \quad i=1, \ldots, n.
\end{align}
Here, $Q_\tau\left(Y_i \mid \boldsymbol{f}_i, \boldsymbol{u}_i\right)$ denotes the conditional $\tau$-quantile of the response $Y_i$ given the latent common factors $\boldsymbol{f}_i \in \mathrm{R}^M$ and the idiosyncratic components $\boldsymbol{u}_i \in\mathrm{R}^{d}$. The vectors $\boldsymbol{\gamma}^*(\tau) \in \mathrm{R}^M$ and $\boldsymbol{\beta}^*(\tau) \in \mathrm{R}^d$ represent the regression coefficients associated with $\boldsymbol{f}_i$ and $\boldsymbol{u}_i$ at the $\tau$-th quantile, respectively. Furthermore, let $\varepsilon_{i}(\tau)=Y_{i}-Q_\tau\left(Y_{i} \mid \boldsymbol{f}_i, \boldsymbol{u}_i\right)$, it is easy to verify that $Q_\tau\left( \varepsilon_{i}(\tau)\mid \boldsymbol{f}_i, \boldsymbol{u}_i\right)=0$. 

Through algebraic operations, we obtain an equivalent representation in terms of the observable covariates $\boldsymbol{X}_{i}$ and the common factors $\boldsymbol{f}_{i}$,
\begin{align}
\label{e3}
Q_\tau\left(Y_{i} \mid \boldsymbol{X}_i\right)=\boldsymbol{f}_i^\top\boldsymbol{\gamma}^*(\tau)+(\boldsymbol{X}_i-\boldsymbol{B}\boldsymbol{f}_{i})^\top\boldsymbol{\beta}^*(\tau)=\boldsymbol{f}_i^\top\boldsymbol{\varphi}^*(\tau)+\boldsymbol{X}_i^\top\boldsymbol{\beta}^*(\tau). 
\end{align}

where $\boldsymbol{\varphi}^*(\tau) := \boldsymbol{\gamma}^*(\tau) - \boldsymbol{B}^\top\boldsymbol{\beta}^*(\tau)$ quantifies the extra contribution of the latent factor $\boldsymbol{f}$ beyond the observed predictor $\boldsymbol{X}$. This extra contribution constitutes a key advantage of the FAQR structure \eqref{e5}. 
% Compared to $Q_\tau\left(Y_{i} \mid \boldsymbol{f}_i, \boldsymbol{u}_i\right)=(\boldsymbol{B} \boldsymbol{f}_i+\boldsymbol{u}_i)^{\top} \boldsymbol{\beta}^{*}(\tau)=\boldsymbol{f}_{i}\boldsymbol{B}^{\top}\boldsymbol{\beta}^{*}(\tau)+\boldsymbol{u}_{i}^{\top}\boldsymbol{\beta}^{*}(\tau)$ obtained by substituting \eqref{e5} into the sparse model $Q_\tau\left(Y_{i} \mid \boldsymbol{f}_i, \boldsymbol{u}_i\right)=\boldsymbol{X}_{i}^{\top} \boldsymbol{\beta}^{*}(\tau)$ , the latent factor $\boldsymbol{f}$ in the model \eqref{e5} has an additional contribution to $Y$. \cite{fan2023latent} argues that in fact, especially when variables are highly correlated, the common factor is likely to have an additional contribution to the response, rather than just a fixed portion $\boldsymbol{B^{\top}\beta^{*}}(\tau)$. 
Throughout this paper, we assume that $M$ is small and remains fixed as suggested by Fan et al.\cite{fan2023latent}.

\subsection{Estimation}
This subsection details the estimation procedure of the FAQR framework. We first estimate common factors and idiosyncratic components through principal component analysis (PCA) (Section \ref{sec:fac}). Subsequently, we formulate a convolution-smoothed $\ell_1$-penalized quantile regression objective function to handle high-dimensional and non-smooth problems using the estimated common factors and idiosyncratic components (Section \ref{sec:cs}). The optimization problem is solved by the I-LAMM algorithm with statistical guarantees (Section \ref{sec:I-LAMM}). Finally, we give a data-driven selection scheme for the regularization parameter $\lambda$ (Section \ref{sec:tun}).

\subsubsection{Factor Estimation}\label{sec:fac}

We reformulate the first part of model \eqref{e5} into a compact matrix representation as
\begin{align} \label{eq4}
\mathbb{X} = \boldsymbol{F}\boldsymbol{B}^\top + \boldsymbol{U},
\end{align} 
where $\mathbb{X} = (\boldsymbol{X}_1, \ldots, \boldsymbol{X}_n)^\top$, $\boldsymbol{F} = (\boldsymbol{f}_1, \ldots, \boldsymbol{f}_n)^\top$, and $\boldsymbol{U} = (\boldsymbol{u}_1, \ldots, \boldsymbol{u}_n)^\top$. Throughout this paper, we assume that we observe $\{(\boldsymbol{X}_i, Y_i)\}_{i=1}^n$, while the common factors $\boldsymbol{F}$ and the idiosyncratic components $\boldsymbol{U}$ are unobserved and need to be estimated from $\mathbb{X}$. 
To address the identifiability issue, we impose the following identifiability assumption.  
\begin{assumption} \label{ap1}
    The covariance matrix of the factor $\operatorname{Cov}(\boldsymbol{f}) = \boldsymbol{I}_M$ and $\boldsymbol{B}^\top \boldsymbol{B}$ is diagonal, where $\boldsymbol{I}_M$ denotes the $M$-dimensional identity matrix.  
\end{assumption}  

Under Assumption \ref{ap1}, we estimate $\boldsymbol{F}$ and $\boldsymbol{B}$ from $\mathbb{X}$ through the principal components analysis (PCA)-based approach, solving
$$
\begin{aligned}
    (\widehat{\boldsymbol{F}}, \widehat{\boldsymbol{B}}) = & \operatorname*{arg\,min}_{\boldsymbol{F} \in \mathbb{R}^{n \times M}, \boldsymbol{B} \in \mathbb{R}^{d \times M}} \|\mathbb{X} - \boldsymbol{F}\boldsymbol{B}^\top\|_{\mathbb{F}}^2, \\
    & \text{subject to } \frac{1}{n} \boldsymbol{F}^\top \boldsymbol{F} = \boldsymbol{I}_M \text{ and } \boldsymbol{B}^\top \boldsymbol{B} \text{ is diagonal}.
\end{aligned}
$$
By Bai \cite{bai2003inferential}, the columns of $\widehat{\boldsymbol{F}}/\sqrt{n}$ correspond to the eigenvectors associated with the $M$ largest eigenvalues of the matrix $\mathbb{X}\mathbb{X}^\top$, and  
\begin{eqnarray}
\widehat{\boldsymbol{B}}^{\top}= (\widehat{\boldsymbol{F}}^\top \widehat{\boldsymbol{F}})^{-1} \widehat{\boldsymbol{F}}^\top \mathbb{X} = n^{-1} \widehat{\boldsymbol{F}}^\top \mathbb{X},~~
\widehat{\boldsymbol{U}}= \mathbb{X} - \widehat{\boldsymbol{F}}\widehat{\boldsymbol{B}}^\top = (\boldsymbol{I}_n - n^{-1} \widehat{\boldsymbol{F}}\widehat{\boldsymbol{F}}^\top) \mathbb{X}.\notag
\end{eqnarray}

In practice, the number of latent factors, $M$, is often unknown and must be determined data-driven. In this study, we adopt the ratio-based method \citep{lam2012factor,ahn2013eigenvalue}. Let $\lambda_m(\mathbb{X} \mathbb{X}^\top)$ denote the $m$-th largest eigenvalue of $\mathbb{X} \mathbb{X}^\top$. The number of factors can be consistently estimated as  
$$
\widehat{M} = \arg \max_{m \leq M_{\max}} \frac{\lambda_m(\mathbb{X} \mathbb{X}^\top)}{\lambda_{m+1}(\mathbb{X} \mathbb{X}^\top)},
$$  
where $1 \leq M_{\max} \leq n$ is a pre-specified upper bound for $M$. In line with Li and Li\cite{li2022integrative}, we treat $M$ as a fixed constant, recognizing that it is typically small due to its connection with the spiked eigenvalues of $\mathbb{X} \mathbb{X}^\top$. For theoretical analyses, we assume $M$ is known; all results remain valid under the condition that $\widehat{M}$ consistently estimates $M$.

\subsubsection{Convolution-type Smoothing Penalized Quantile Regression}\label{sec:cs}

With estimated common factors and idiosyncratic components, we proceed to estimate the regression parameters in the FAQR model. Let $\widehat{\boldsymbol{Z}}_i = (\widehat{\boldsymbol{u}}_i^\top, \widehat{\boldsymbol{f}}_i^\top)^\top$ be the $(d+M)$-vector formed by stacking the estimated idiosyncratic components and common factors for the subject $i$. In high-dimensional settings, where the number of predictors $d$ can far exceed the sample size $n$, it is generally assumed that only a subset of the predictors significantly influences the response variable. This assumption introduces sparsity in the true parameter vector $\boldsymbol{\beta}^*(\tau)$. Exploiting this sparsity allows for more efficient and interpretable parameter estimation in high-dimensional models. 

To achieve sparse estimation, a common approach is to apply $\ell_1$-penalty \citep{10.1214/10-AOS827}, that is
\begin{align}\label{eq3}
\widetilde{\boldsymbol{\theta}}(\tau)=\underset{\boldsymbol{\theta} \in \mathbb{R}^{d+M}}{\operatorname{argmin}} \Bigg\{ 
    \underbrace{\frac{1}{n} \sum_{i=1}^n \rho_\tau\left(Y_i - \widehat{\boldsymbol{Z}}_i^{\top} \boldsymbol{\theta}\right)}_{=: \widehat{Q}(\boldsymbol{\theta};\tau)} 
    + \lambda\|\widehat{\boldsymbol{\sigma}}\odot\boldsymbol{\theta}\|_1
    \Bigg\},
\end{align}
where $\boldsymbol{\theta} = (\boldsymbol{\beta}^{\top},\boldsymbol{\gamma}^{\top})^{\top}$, $\rho_\tau(t) = t\{\tau - \mathbb{I}(t < 0)\}$ represents the quantile loss function at level $\tau$, $\widehat{\boldsymbol{\sigma}}=(\widehat{\sigma}_1,\dots,\widehat{\sigma}_{d+M})^\top$, $\widehat{\sigma}_j^2$ is the empirical variance of $j$-th component of $\widehat{\boldsymbol{Z}}_i$, and $\odot$ is the Hadamard product. The term $\lambda \|\cdot\|_1$ denotes the $\ell_1$-penalty, where $\lambda \ge 0$ is a tuning parameter.

One key challenge of the quantile loss function $\rho_\tau(\cdot)$ lies in its lack of differentiability, which complicates optimization, particularly in high-dimensional settings where the sample covariance matrix of $\widehat{\boldsymbol{Z}}$ may be singular. To overcome this, we adopt the convolution-smoothed quantile loss function \citep{fernandes2021smoothing, tan2022high}, approximating $\widehat{Q}(\boldsymbol{\theta};\tau)$ by 
$$
\widehat{Q}_h(\boldsymbol{\theta};\tau)=\int_{-\infty}^{\infty} \rho_\tau(t) \mathrm{d} \widehat{G}_h(t ; \boldsymbol{\theta})=\frac{1}{n h} \sum_{i=1}^n \int_{-\infty}^{\infty} \rho_\tau(t) K\left(\frac{t-r_i(\boldsymbol{\theta})}{h}\right) \mathrm{d} t,
$$ 
with $r_i(\boldsymbol{\theta})=Y_i - \widehat{\boldsymbol{Z}}_i^{\top} \boldsymbol{\theta}$,
\begin{eqnarray}
\widehat{G}_h(t; \boldsymbol{\theta}) &=& \int_{-\infty}^t \widehat{g}_h(s; \boldsymbol{\theta}) \, \mathrm{d}s,\notag\\
\widehat{g}_h(t; \boldsymbol{\theta}) &=& \frac{1}{n} \sum_{i=1}^n K_h\left(t - r_i(\boldsymbol{\theta})\right), \quad K_h(t) = \frac{1}{h} K\left(\frac{t}{h}\right),
\end{eqnarray} 
$K:\mathbb{R} \rightarrow [0, \infty)$ being a nonnegative symmetric kernel function with the bandwidth $h$, such as a Gaussian kernel or a Laplacian kernel.

Then, we estimate $\boldsymbol{\theta}$ by the $\ell_1$-penalized smoothed quantile regression ($\ell_1$-SQR) estimator,
\begin{align}\label{eq:sqr}
    \widehat{\boldsymbol{\theta}}_h(\tau) = \underset{\boldsymbol{\theta} \in \mathbb{R}^{d+M}}{\operatorname{argmin}} \left\{ \widehat{Q}_h(\boldsymbol{\theta};\tau) + \lambda \|\widehat{\boldsymbol{\sigma}}\odot\boldsymbol{\theta}\|_1 \right\}.
\end{align}

By choosing a sufficiently smooth kernel function $K(\cdot)$, the smoothed loss function $\widehat{Q}_h(\boldsymbol{\theta};\tau)$ become twice continuously differentiable, and
the gradient and the Hessian matrix of $\widehat{Q}_h(\boldsymbol{\theta};\tau)$ with respect to $\boldsymbol{\theta}$ are respectively   
$$
\begin{aligned}
\nabla_{\boldsymbol{\theta}} \widehat{Q}_h(\boldsymbol{\theta};\tau) &= \frac{1}{n} \sum_{i=1}^n \left[\bar{K}_h\left(-r_i(\boldsymbol{\theta})\right) - \tau\right] \widehat{\boldsymbol{Z}}_i, \\
\nabla_{\boldsymbol{\theta}}^2 \widehat{Q}_h(\boldsymbol{\theta};\tau) &= \frac{1}{n} \sum_{i=1}^n K_h\left(-r_i(\boldsymbol{\theta})\right) \widehat{\boldsymbol{Z}}_i \widehat{\boldsymbol{Z}}_i^\top,
\end{aligned}
$$  
where $\bar{K}(t) = \int_{-\infty}^t K(s) \, \mathrm{d}s$ and $\bar{K}_h(t) = \bar{K}(t/h)$.  

This approach effectively combines the benefits of regularized optimization and convolution smoothing to handle high-dimensional data with sparsity constraints and non-smooth loss function of quantile regression. For notational ease, we simplify $\widehat{Q}_h(\boldsymbol{\theta};\tau)$ to $\widehat{Q}_h(\boldsymbol{\theta})$, and
$\widehat{\boldsymbol{\theta}}_h(\tau)$ to $\widehat{\boldsymbol{\theta}}$ in the subsequent sections.

\subsubsection{Implementation via I-LAMM Algorithm}\label{sec:I-LAMM}

To solve the $\ell_1$-penalized convolution smoothed quantile regression problem efficiently, we adopt the I-LAMM algorithm \citep{fan2018lamm, tan2022high}, which solves the optimization problem in polynomial time and achieves optimal statistical accuracy.

Recall that 
$$
\widehat{Q}_h(\boldsymbol{\theta}) = \frac{1}{nh} \sum_{i=1}^n \int_{-\infty}^\infty \rho_\tau(t) K\left(\frac{t + \widehat{\boldsymbol{Z}}_i^\top \boldsymbol{\theta} - Y_i}{h}\right) \, \mathrm{d}t.  
$$
Given an initial estimate $\widetilde{\boldsymbol{\theta}}^{(0)}$ at iteration 0, we iteratively solve an isotropic quadratic approximation to $\widehat{Q}_h(\boldsymbol{\theta})$. At iteration $\ell$, the approximation is  
$$
\mathcal{L}(\boldsymbol{\theta}; \tilde{\boldsymbol{\theta}}^{(\ell-1)}, \phi) = \widehat{Q}_h(\tilde{\boldsymbol{\theta}}^{(\ell-1)}) + \langle \nabla_{\boldsymbol{\theta}} \widehat{Q}_h(\tilde{\boldsymbol{\theta}}^{(\ell-1)}), \boldsymbol{\theta} - \tilde{\boldsymbol{\theta}}^{(\ell-1)} \rangle + \frac{\phi}{2} \|\boldsymbol{\theta} - \tilde{\boldsymbol{\theta}}^{(\ell-1)}\|_2^2,  
$$  
where $\phi$ is a parameter controlling the quadratic term, which is at least as large as the largest eigenvalue of $\nabla_{\boldsymbol{\theta}}^2 \widehat{Q}_h(\tilde{\boldsymbol{\theta}}^{(\ell-1)})$ based on Taylor's expansion. This isotropic form enables an analytic solution to the optimization problem  
$$
\underset{\boldsymbol{\theta} \in \mathbb{R}^{d+M}}{\operatorname{argmin}} \Bigg\{\widehat{Q}_h(\tilde{\boldsymbol{\theta}}^{(\ell-1)}) + \langle \nabla_{\boldsymbol{\theta}} \widehat{Q}_h(\tilde{\boldsymbol{\theta}}^{(\ell-1)}), \boldsymbol{\theta} - \tilde{\boldsymbol{\theta}}^{(\ell-1)} \rangle + \frac{\phi}{2} \|\boldsymbol{\theta} - \tilde{\boldsymbol{\theta}}^{(\ell-1)}\|_2^2 + \lambda \|\widehat{\boldsymbol{\sigma}}\odot\boldsymbol{\theta}\|_1 \Bigg\},
$$  
with minimizer  
$$
\tilde{\boldsymbol{\theta}}^{(\ell)} = T_\lambda\left(\tilde{\boldsymbol{\theta}}^{(\ell-1)}; \phi\right) = S\left(\tilde{\boldsymbol{\theta}}^{(\ell-1)} - \phi^{-1} \nabla_{\boldsymbol{\theta}}\widehat{Q}_h(\tilde{\boldsymbol{\theta}}^{(\ell-1)}); \phi^{-1} \lambda\widehat{\boldsymbol{\sigma}}\right),  
$$  
where $S(\cdot;\cdot)$ is the component-wise soft-thresholding operator.

In practice, computing eigenvalues of $\nabla_{\boldsymbol{\theta}}^2 \widehat{Q}_h(\tilde{\boldsymbol{\theta}}^{(\ell-1)})$ can be challenging. Instead, we can increase the value of the function $\mathcal{L}(\cdot;\cdot,\cdot)$ little by little until it exceeds $\widehat{Q}_{h}(\cdot)$. To achieve this, we adjust $\phi$ iteratively. Starting from a small initial value $\phi_0$, we update it by multiplying by a factor $c_0 > 1$ at each iteration,
$$
\phi^{(\ell, k)} = c_0^{k-1} \phi_0,  
$$  
where $k$ denotes the iteration index. Using this $\phi^{(\ell, k)}$, we get the updated parameter $\tilde{\boldsymbol{\theta}}^{(\ell, k)}$ as  
$$
\tilde{\boldsymbol{\theta}}^{(\ell, k)} = T_\lambda\left(\tilde{\boldsymbol{\theta}}^{(\ell, k-1)}; \phi^{(\ell, k)}\right),  
$$  
with $\tilde{\boldsymbol{\theta}}^{(\ell, k-1)} = \tilde{\boldsymbol{\theta}}^{(\ell-1)}$. This process continues until the condition $\mathcal{L}(\cdot;\cdot,\cdot) > \widehat{Q}_h(\cdot)$ is met. Algorithm \ref{al:1} shows this process.

\begin{algorithm}[t]
    \caption{}
    \label{al:1}
        \begin{enumerate}
        \item \textbf{Input}: Data, $\widetilde{\boldsymbol{\theta}}^{(0)}, \phi_0,\varepsilon$ (tolerance parameter)
        \item \textbf{For $\ell=1,2,\cdots$}: $\phi = \phi_{0}$ 
        \begin{enumerate}
            \item  
            \begin{itemize}
            \item $k=1$, $\phi^{(\ell,1)}=\phi_{0}$ and  $\tilde{\boldsymbol{\theta}}^{(\ell,0})=\tilde{\boldsymbol{\theta}}^{(\ell-1)}$
            \item $\tilde{\boldsymbol{\theta}}^{(\ell, 1)} = T_{\lambda}(\tilde{\boldsymbol{\theta}}^{(\ell, 0)}; \phi^{(\ell, 1)})$
            \item \textbf{If} 
        $$
            \begin{aligned}
                \mathcal{L}_{\lambda}\left(\tilde{\boldsymbol{\theta}}^{(\ell, 1)} ; \tilde{\boldsymbol{\theta}}^{(\ell, 0)},\phi^{(\ell, 0)}\right)\geq \widehat{Q}_{h}\left(\tilde{\boldsymbol{\theta}}^{(\ell, 1)}\right)
            \end{aligned}
            $$
            \item Then $\phi=\phi_{0}$
            \item \textbf{Else} for $k=1,2,\cdots$:  
            $$
            \begin{aligned}
               &\phi^{(\ell, k)} \leftarrow  c_0^{k-1} \phi_0,  \quad\tilde{\boldsymbol{\theta}}^{(\ell, k)} \leftarrow T_{\lambda}\left(\tilde{\boldsymbol{\theta}}^{(\ell, k-1)};\phi^{(\ell-1, k)}\right)\\
               &k=k+1
            \end{aligned}
            $$
            \item Until $\mathcal{L}_{\lambda}\left(\tilde{\boldsymbol{\theta}}^{(\ell, k)} ; \tilde{\boldsymbol{\theta}}^{(\ell, k-1)},\phi^{(\ell,k)}\right)\geq \widehat{Q}_{h}\left(\tilde{\boldsymbol{\theta}}^{(\ell, k)}\right)$ 
        \end{itemize}
        \item \textbf{Return} $\left\{\tilde{\boldsymbol{\theta}}^{(\ell, k)}, \phi^{(\ell, k)}\right\}$           
        \end{enumerate}
        \item 
            $\tilde{\boldsymbol{\theta}}^{(l)}=\tilde{\boldsymbol{\theta}}^{(\ell, k)}$\\
            $\ell=\ell+1$
        \item \textbf{Repeat} step 2 and 3 \textbf{Until} $\|\tilde{\boldsymbol{\theta}}^{(\ell)}-\tilde{\boldsymbol{\theta}}^{(\ell-1)}\|<\varepsilon$
        \item \textbf{Return} $\widehat{\boldsymbol{\theta}}=\tilde{\boldsymbol{\theta}}^{(\ell)}$
    \end{enumerate}
\end{algorithm}

Rather than setting $\widetilde{\boldsymbol{\theta}}^{(0)}=\boldsymbol{0}$, we give a warm start by the robust expectile regression estimator as the code in Tan et al. \cite{tan2022high},

\begin{align}\label{eq:ws}
\widetilde{\boldsymbol{\theta}}^{(0)}=\underset{\boldsymbol{\theta} \in \mathbb{R}^{d+M}}{\operatorname{argmin}}\left\{\sum_{i=1}^n H_c\left(Y_i-\widehat{\boldsymbol{Z}}_i^{\top} \boldsymbol{\theta}\right) *\mid\tau-I\left(Y_i-\widehat{\boldsymbol{Z}}_i^{\top} \boldsymbol{\theta}<0\right)\mid+\lambda\|\widehat{\boldsymbol{\sigma}} \odot \boldsymbol{\theta}\|_1\right\},    
\end{align}
where
$$
H_c(z)=\left\{\begin{array}{cc}
z^2 / 2, & \text { if } \mid z\mid \leqslant c, \\
c z-c^2 / 2, & \text { if }\mid z\mid>c,
\end{array}\right.
$$
and $c>0$ is the robustness parameter which balances robustness and bias, being specified later in the numerical study. 

\subsubsection{Choice of $\lambda$}\label{sec:tun}
We select $\lambda$ following the seminal work of Belloni and Chernozhukov \cite{10.1214/10-AOS827} who established a data-driven selection scheme for the penalization parameters. First, define the pivotal quantity
$$
\Lambda = \max_{1 \leq j \leq d+M} \mid \sum_{i=1}^n \frac{\widehat{z}_{ij} \{\tau - \mathbb{I}(e_i \leq \tau)\}}{\widehat{\sigma}_j} \mid,
$$
where $\widehat{z}_{ij}$ is $j$-th component of $\widehat{\boldsymbol{Z}}_i$ and $\{e_i\}_{i=1}^n$ are independently uniformly distributed error terms on (0,1). Then the regularization parameter is determined by
$$
\lambda = c_0 \cdot Q_{1-\alpha}(\Lambda \mid \mathcal{Z}),
$$
where $\mathcal{Z}=\{\widehat{\boldsymbol{Z}}_i,i=1,\dots,n\}$, $c_0>1$ denotes a scaling constant, $Q_{1-\alpha}(\Lambda \mid \mathcal{Z})$ indicates the empirical $(1-\alpha)$-quantile of $\Lambda$; we set $c_0=1.1$ and $\alpha=0.1$ in numerical studies.

\section{Asymptotic Properties}

We now proceed to establish the theoretical properties of the proposed $\ell_1$-SQR estimator $\widehat{\boldsymbol{\theta}}$, including the consistency and convergence rates. The following additional assumptions are needed for the theoretical derivations.

\begin{assumption}[Conditions on the factor model]\label{assump:regularity}
\begin{enumerate}
    \item[(a)] There exists a positive constant $C_1<\infty$ such that $\|\boldsymbol{f}\|_{\psi_2} \leqslant C_1$ and $\|\boldsymbol{u}\|_{\psi_2} \leqslant C_1$.
    \item[(b)] There exists a constant $C_{2}>1$ such that $d / C_{2} \leqslant \lambda_{\min }\left(\boldsymbol{B}^{\top} \boldsymbol{B}\right) \leqslant \lambda_{\max }\left(\boldsymbol{B}^{\top} \boldsymbol{B}\right) \leqslant d C_{2}$.

    \item[(c)] Let $\boldsymbol{\Sigma}=\operatorname{Cov}(\boldsymbol{u})$. There exists a constant $C_{3}>0$ such that $\|\boldsymbol{B}\|_{\max } \leqslant C_{3}$ and
$$
\mathbb{E}\mid\boldsymbol{u}^{\top} \boldsymbol{u}-\operatorname{tr}(\boldsymbol{\Sigma})\mid^4 \leqslant C_{3} d^2.
$$

    \item[(d)] There exist a positive constant $C_{4}<1$ such that $C_{4} \leqslant \lambda_{\min }(\boldsymbol{\Sigma}),\|\boldsymbol{\Sigma}\|_1 \leqslant 1 / C_{4}$ and $\min _{1 \leqslant k, \ell \leqslant d} \operatorname{Var}\left(u_k u_{\ell}\right) \geqslant C_{4}$.
\end{enumerate}
     
\end{assumption}

\begin{assumption}[Sparsity condition]\label{assump:sparsity}
The true parameter vector $\boldsymbol{\theta}^*$ is $s$-sparse, which is $\left\|\boldsymbol{\theta}^*\right\|_0 \leq s$, and we assume that
$$
s \cdot\left(\frac{\log d \log n}{n}\right)^{1 / 5}=o(1).
$$
\end{assumption}

\begin{assumption}[Conditions on the kernel function and noise density]\label{ap4}  
\begin{enumerate}
\item[(a)] The kernel function $K(\cdot)$ is non-negative, symmetric, and satisfies the normalization condition $\int_{-\infty}^\infty K(t) \, \mathrm{d}t = 1$. It is continuously differentiable up to the second order, with its zero-order, first-order, and second-order derivatives uniformly bounded. Specifically, we assume:  
$$
\overline{\zeta} = \sup_{t \in \mathbb{R}} K(t) < \infty, \underline{\zeta} = \min_{t \in [-1, 1]} K(t) > 0,
$$
$$
\overline{\zeta}^\prime = \sup_{t \in \mathbb{R}} \mid K^\prime(t)\mid < \infty, \overline{\zeta}^{\prime\prime} = \sup_{t \in \mathbb{R}} \mid K^{\prime\prime}(t)\mid < \infty.
$$  
Additionally, define $\zeta_k = \int_{-\infty}^\infty |t|^k K(t) \, \mathrm{d}t$ for $k \geq 1$.  
\label{ap2}

\item[(b)] The conditional density function $g_{\varepsilon \mid \boldsymbol{Z}}(\cdot)$ satisfies a Lipschitz condition with constant $L_0 > 0$, that is $\mid g_{\varepsilon \mid \boldsymbol{Z}}(t_1) - g_{\varepsilon \mid \boldsymbol{Z}}(t_2)\mid  \leq L_0 \mid t_1 - t_2\mid , \quad \forall t_1, t_2 \in \mathbb{R},$ almost surely over $\boldsymbol{Z}$. Furthermore, there exist constants $\bar{g} \geq \underline{g} > 0$ such that $g_{\varepsilon \mid \boldsymbol{Z}}(0) \geq \underline{g}$ and $\sup_{t \in \mathbb{R}} g_{\varepsilon \mid \boldsymbol{Z}}(t) \leq \bar{g}$.
\label{ap3}
\end{enumerate}  
\end{assumption}  
Assumptions \ref{assump:regularity} (a)-(d) are standard in the analysis of high-dimensional factor models. For further details, we refer to \cite{bai2003inferential}, \cite{fan2013large}, and \cite{li2018embracing}.
Assumption \ref{assump:sparsity} is standard in the high-dimensional literature.
Assumption \ref{ap4}(a) imposes standard properties on the kernel function, including additional constraints on its derivatives. The condition $\underline{\zeta} = \min_{t \in [-1, 1]} K(t) > 0$ is introduced for theoretical simplicity and can be generalized to $\min_{t \in [-c, c]} K(t) > 0$ for some $c \in (0, 1)$. Widely used kernel functions such as the Gaussian kernel, the Epanechnikov kernel, and the logistic kernel, along with their rescaled versions, satisfy these conditions.  
Assumption \ref{ap4}(b) establishes regularity conditions for the conditional density function. These conditions are standard in quantile regression analysis and provide the foundation for deriving theoretical properties of the smoothed quantile regression estimator.

Given the above assumptions, we have the following result.  
\begin{theorem}\label{Th1}
Suppose Assumptions \ref{ap1}-\ref{ap4} hold. Under FAQR \eqref{e5} with $\boldsymbol{\theta}^{*}$ being $s$-sparse, the $\ell_1$-SQR estimator $\widehat{\boldsymbol{\theta}}$, with $\lambda \asymp s\sigma  \sqrt{\frac{\log d \log n}{n}}$, satisfies
$$
\left\|\widehat{\boldsymbol{\theta}} - \boldsymbol{\theta}^*\right\|_2 = O_p(s^{1/2} \lambda) \quad \text{and} \quad \left\|\widehat{\boldsymbol{\theta}} - \boldsymbol{\theta}^*\right\|_1 = O_p(s \lambda),
$$
provided that the bandwidth $h$ satisfies 
$$
\max \left(\frac{\sigma}{\underline{g}} \sqrt{\frac{s \log d}{n}}, \frac{\sigma \bar{g}}{\underline{g}^2} \frac{s \log d}{n}\right) \lesssim h \leq \min \left\{\frac{\underline{g}}{2L_0}, \sqrt{s^{1/2} \lambda}\right\},
$$  
where $\sigma^2=\max\{\max_{1\leq j\leq d}E(u_j^2),\max_{1\leq j\leq M}E(f_j^2)\}$.
\end{theorem}  

\textbf{Remark 3.1} (Tuning selection in numerical implementation). In our numerical experiments, the regularization parameter $\lambda$ was selected using the data-driven scheme proposed by Belloni and Chernozhukov \cite{10.1214/10-AOS827}, which is designed to satisfy the scaling requirement $\lambda \asymp s \sigma \sqrt{\log d \log n / n}$ inherent in Theorem \ref{Th1} under appropriate conditions. The bandwidth $h$ was set to $\max\{0.05, \tau(1-\tau)[\log (d+M)/n]^{1/4}\}$ for the smoothed quantile loss. This choice adheres to the theoretical constraints specified in Theorem \ref{Th1}: the lower bound (0.05) prevents excessive smoothing for finite samples, while the term $[\log d/n]^{1/4}$ ensures the asymptotic decay rate of $h$ is compatible with the upper bound requirement and the convergence rate conditions stipulated in the theorem. Consequently, the tuning parameters in our simulations are consistent with the theoretical framework established above.

Theorem \ref{Th1} establishes error bounds for the $\ell_1$-SQR estimator $\widehat{\boldsymbol{\theta}}$ in both $\ell_1$- and $\ell_2$-norms. These results demonstrate that the estimator achieves consistency under high-dimensional settings. Notably, the derived bounds hold without requiring the covariates $\boldsymbol{X}$ to exhibit bounded sub-Gaussian norms, a common assumption in traditional high-dimensional analysis. This relaxation aligns with the frameworks presented in recent studies, such as He et al. \cite{he2023smoothed} and Yan et al. \cite{yan2023confidence}.  

The results highlight the efficacy of the smoothing approach, particularly in addressing the challenges posed by high-dimensional quantile regression settings. The interplay between the tuning parameter $\lambda$ and the bandwidth $h$ ensures that the estimator remains robust while maintaining desirable statistical properties.

\section{Adequacy Test of Factor Model}
The latent factor regression is widely applied in many fields as an efficient dimension reduction method. A natural question arises is whether the model is adequate and FAQR (\ref{e5}) serves naturally as the alternative model. To be more specific, we consider testing the hypotheses

\begin{align}
    \label{eq23}
    H_0: \boldsymbol{\beta}^{*}=0 \text { versus } H_1: \boldsymbol{\beta}^{*} \neq 0
\end{align}
in FAQR (\ref{e5}).

Following the idea of Tang et al. \cite{tang2022conditional}, we propose a rescaled conditional marginal score statistic. To detect the significance of $\widehat{\boldsymbol{u}}$ in the presence of $\widehat{\boldsymbol{f}}$, we construct a score-type test statistic as follows.

First, we estimate the marginal effect of $\widehat{\boldsymbol{f}}$ as
$$
\widehat{\boldsymbol{\gamma}}=\underset{\boldsymbol{\gamma} \in \mathbb{R}^M}{\operatorname{argmin}}\quad \widehat{Q}_h(\boldsymbol{\gamma}),
$$
where 
$$
\widehat{Q}_h(\boldsymbol{\gamma}):=\int_{-\infty}^{\infty} \rho_\tau(t) \mathrm{d} \hat{G}_h(t ; \boldsymbol{\gamma})=\frac{1}{n h} \sum_{i=1}^n \int_{-\infty}^{\infty} \rho_\tau(t) K\left(\frac{t+\widehat{\boldsymbol{f}}_i^{\top} \boldsymbol{\gamma}-Y_i}{h}\right) \mathrm{d} t .
$$
To evaluate the additional effect of each $\widehat{u}_{j}$ conditional on $\widehat{\boldsymbol{f}}$, we project $\widehat{u}_{j}$ on $\widehat{\boldsymbol{f}}$ with weights $\widehat{\mathbf{K}}_\tau$ to obtain
$$
\widehat{\boldsymbol{u}}^{*}_{\cdot j}=\left\{\mathbb{I}_n-\widehat{\mathbf{K}}_\tau \widehat{\boldsymbol{F}}\left(\widehat{\boldsymbol{F}}^{\top} \widehat{\mathbf{K}}_\tau^2 \widehat{\boldsymbol{F}}\right)^{-1} \widehat{\boldsymbol{F}}^{\top} \widehat{\mathbf{K}}_\tau\right\} \widehat{\boldsymbol{u}}_{\cdot j}
$$
where $\widehat{\boldsymbol{u}}_{\cdot j}=(\widehat{u}_{1j},\dots,\widehat{u}_{nj})^\top$, $\widehat{\mathbf{K}}_\tau=\operatorname{diag}\left\{K_{h}\left(-\varepsilon_{1}\right),\cdots,K_{h}\left(-\varepsilon_{2}\right)\right\}$ such that the $j$ th component of $\boldsymbol{u}$ is orthogonal to $\widehat{\boldsymbol{f}}$ in a weighted manner; that is, $\widehat{\boldsymbol{F}}^{\top} \widehat{\mathbf{K}}_\tau \widehat{\boldsymbol{u}}^{*}_{\cdot j}=\mathbf{0}$, for $j=1, \ldots, d$. 

% We consider the weighted projection to account for the heteroscedasticity using $f_{i, \tau}(\cdot)$, thus eliminating the first-order difference; see the proof of Theorem 1 in the Supplementary Material (equation (S.16), Section S3.2) for more details. Similar projections can also be found in the quantile literature; see for instance, Park and He (2017).

Second, we define the rescaled conditional marginal score statistic as
$$
S_{\tau}\left(\boldsymbol{\gamma};
\widehat{\boldsymbol{u}},\widehat{\boldsymbol{f}}\right)=\frac{1}{n} \sum_{i=1}^n \left[ \bar{K}_h(-r_i(\boldsymbol{\gamma}))  -\tau \right]\widehat{\boldsymbol{u}}^{*}_{i}.
$$

Finally, the proposed maximum-score test statistic is defined as
$$
T_{n}(\tau)= \|S_{\tau}(\widehat{\boldsymbol{\gamma}};
\widehat{\boldsymbol{u}},\widehat{\boldsymbol{f}})\|_{\infty}.
$$

\begin{theorem} \label{th2}
    Suppose that conditions $\mathrm{A} 1-\mathrm{A} 3$ hold. Then, we have
    $$
    \|S_{\tau}(\widehat{\boldsymbol{\gamma}};\widehat{\boldsymbol{u}},\widehat{\boldsymbol{f}})-S_{\tau}\left(\boldsymbol{\gamma}^{*};\boldsymbol{u},\boldsymbol{f}\right)\|_{\infty}=O_p \left(s\left\{\sqrt{\frac{\log n}{d}}+\sqrt{\frac{(\log d)(\log n)}{n}}\right\}\right)
    $$
as $n, d\rightarrow \infty$, where $\sqrt{n}S_{\tau}\left(\boldsymbol{\gamma}^{*};\boldsymbol{u},\boldsymbol{f}\right)=\frac{1}{\sqrt{n}} \sum_{i=1}^n \left[ \bar{K}_h(-r_i(\boldsymbol{\gamma}^{*}))  -\tau \right]\boldsymbol{u}^{*}_{i}$ is asymptotic Gaussian.
\end{theorem}
Theorem \ref{th2} provides some convergence result of the proposed test statistic. However, the convergence rate is not satisfactory so that we can not use it for asymptotic calibration. We provide two bootstrap methods for the critical value of $T_{n}(\tau)$.

\textbf{Multiplier Bootstrap}. 
We outline the procedures of the Gaussian multiplier bootstrap in this subsection.
\begin{enumerate}
\item  Let
$$
T_{n,1}^*(\tau)=\left\|\left\{\frac{1}{n} \sum_{i=1}^n w_i \{\tau-\bar{K}_{h}(e_{i})\}\widehat{\boldsymbol{u}}_{i}^{*}\right\}^2 \right\|_{\infty},
$$

where $\left\{e_i ; i=1, \ldots, n\right\}$ is a random sample with the $\tau$ th quantile zero, and $\left\{w_i ; i=1, \ldots, n\right\}$ is a random sample independent of $e_i$ with zero mean, unit variance, and a finite third moment. Specifically, we generate $e_i$ from $N\left(-\Phi^{-1}(\tau), 1\right)$, and $w_i$ from a two-point distribution with $P(w=1)=P(w=-1)=1/2$.

\item  Repeat Step $1$ B times to obtain the bootstrap statistics $\left\{T^{*}_{n, 1}(\tau)^{1}, \ldots\right.$, $\left.T^{*}_{n, 1}(\tau)^{B}\right\}$, and calculate the $P$-value as $B^{-1} \sum_{b=1}^B I\left\{T^{*}_{n, 1}(\tau)^{b}>T_{n, 1}(\tau)\right\}$.
\end{enumerate}

\textbf{Residual Bootstrap}. 
Another possibility is the residual bootstrap. Let $\widehat{\varepsilon}_i(\tau)=Y_i-\hat{\boldsymbol{u}}_i^{\top} \widehat{\boldsymbol{\beta}}(\tau)-\hat{\boldsymbol{f}}_i^{\top} \widehat{\boldsymbol{\gamma}}(\tau)$ be the residuals from the alternative FAQR model. For $b=1,\cdots,B$, repeat the next three steps.
\begin{enumerate}
\item Resample with replacement and equal probability on $\left\{\widehat{\varepsilon}_i(\tau), i=1,2, \cdots, n\right\}$ to obtain the bootstrap sample $\varepsilon_i^{b}(\tau), i=1,2, \cdots, n$, and define $Y_i^{b}=\hat{\boldsymbol{f}}_i^{\top} \widehat{\boldsymbol{\gamma}}(\tau)+\varepsilon_i^{b}(\tau)$.
\item Based on the bootstrap sample $\{Y_{i}^{b},\widehat{\boldsymbol{f}}_{i}\}_{i=1}^{n}$, estimate $\widehat{\boldsymbol{\gamma}}^{b}=\underset{\boldsymbol{\gamma} \in \mathbb{R}^M}{\operatorname{argmin}}\quad \widehat{Q}_h(\boldsymbol{\gamma})$.
\item Compute $T_{n,2}^{b}=\|\boldsymbol{S}_{n}^{b}\|_{\infty}$, where 
$$
S_{n}^{b}=\frac{1}{n}\sum_{i=1}^{n}\{\tau-\bar{K}_{h}(\widehat{\boldsymbol{f}}_{i}^{\top}\widehat{\boldsymbol{\gamma}}^{b}-Y_{i}^{b})\}\widehat{\boldsymbol{u}}_{i}.
$$
\end{enumerate}
The $P$-value is then calculated as $B^{-1} \sum_{b=1}^B I\left\{T_{n,2}^{b}>T_{n, 1}(\tau)\right\}$.

\section{Simulation}
\subsection{Accuracy of Estimation}
We generate data from model (\ref{e5}). We let the number of factors $M=2$, dimension of covariate $d=200, \boldsymbol{\gamma}^{*}=$ $(0.5,0.5)^\top$, the first 3 entries of $\boldsymbol{\beta}^{*}$ be $(1.8,1.6,-1.2)$ and its remaining $d-3$ entries be 0. Throughout this subsection, we generate every entry of $\boldsymbol{F}, \boldsymbol{U}$ from the standard Gaussian distribution;  each entry of $\boldsymbol{B}$ is generated from the uniform distribution $\operatorname{Unif}(-1,1)$, and fixed through replications in each setting. We consider three distributions for the noise: (i) normal distribution $N(0,0.5^2)$, (ii) $t_3$ distribution, and (iii) $t_2$ distribution, respectively. We consider $n=200, 500, 1000$ and $d=200, 500$.

For numerical implementation, we choose a warm start for the optimization problem, obtaining $\widetilde{\boldsymbol{\theta}}^{(0)}$ with loss \eqref{eq:ws}, where
$$
H_c(z)=\left\{\begin{array}{cl}
z^2 / 2, & \text{if}~abs(z) \leqslant c, \\
c z-c^2 / 2, & \text{if}~abs(z)>c.
\end{array}\right.
$$ 
The cutoff parameter is $c = 5\times \min\{\operatorname{mad}(z),\operatorname{sd}(z)\}$, where $\operatorname{mad}(z)$ is the median absolute deviation of $z$ and $\operatorname{sd}(z)$ is the standard deviation. Numerical experiments of Tan et al. \cite{tan2022high} show that the results are rather insensitive to the choice of the bandwidths. The default value of $h$ is set at $\max \left\{0.05, \sqrt{\tau(1-\tau)}\{\log (d+M) / n\}^{1 / 4}\right\}$.

\begin{figure}[h]
    \centering
    \includegraphics[width=0.85\linewidth]{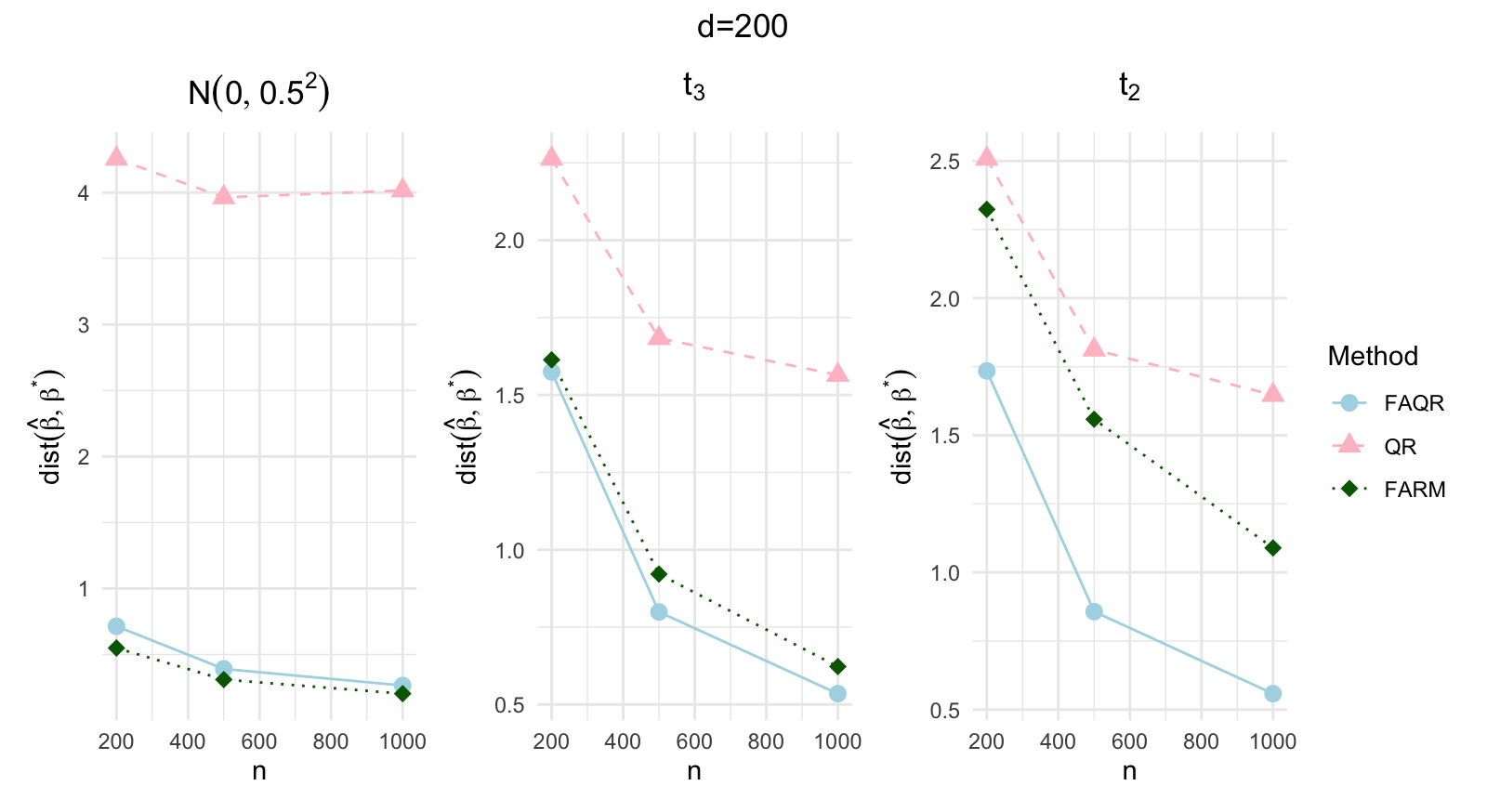}
    \caption{Accuracy for $\boldsymbol{\widehat{\beta}}$ with $\operatorname{dist}\left(\boldsymbol{\widehat{\beta}}, \boldsymbol{\beta}^{*}\right):=\left\|\boldsymbol{\widehat{\beta}}-\boldsymbol{\beta}^{*}\right\|_1$ based on 500 replications with $d=200$ and different noise. FAQR: solid line; FARM: dotted line; QR: dashed line.}
    \label{fig1}
\end{figure}

\begin{figure}[h]
    \centering
    \includegraphics[width=0.85\linewidth]{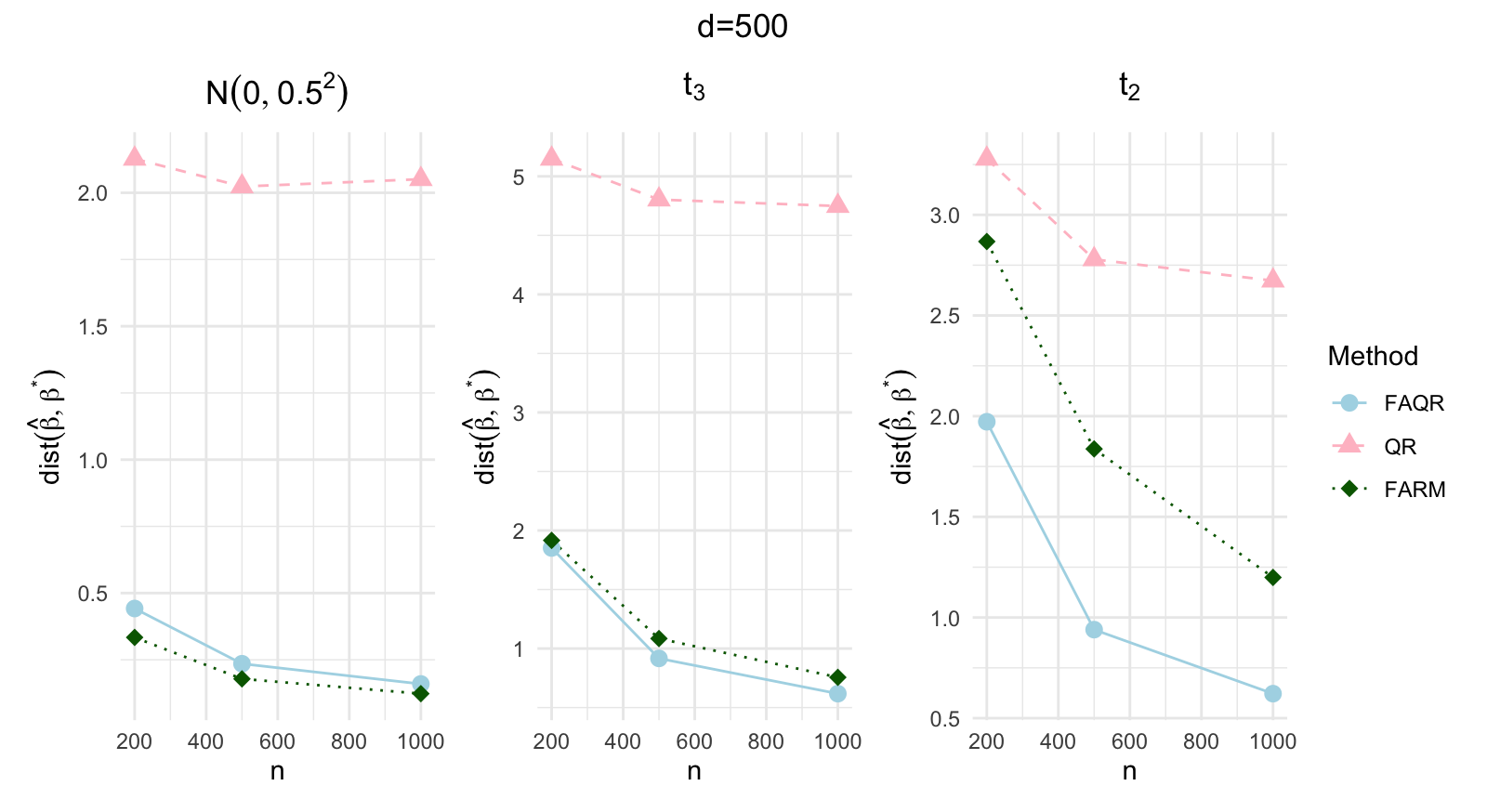}
    \caption{Accuracy for $\boldsymbol{\widehat{\beta}}$ with $\operatorname{dist}\left(\boldsymbol{\widehat{\beta}}, \boldsymbol{\beta}^{*}\right):=\left\|\boldsymbol{\widehat{\beta}}-\boldsymbol{\beta}^{*}\right\|_1$ based on 500 replications with $d=500$ and different noise. FAQR: solid line; FARM: dotted line; QR: dashed line.}
    \label{fig2}
\end{figure}

We compare the proposed FAQR to another two existing methods: (i) QR, convolution-smoothed QR, regressing $Y$ on $\boldsymbol{X}$ directly; (ii) FARM, method in Fan et al. \cite{fan2023latent}. 
Though we jointly estimate $\boldsymbol{\theta} = (\boldsymbol{\gamma}^{\top}, \boldsymbol{\beta}^{\top})^{\top}$, we are interested in the estimation accuracy of $\boldsymbol{\beta}$, presenting $\operatorname{dist}\left(\widehat{\boldsymbol{\beta}}, \boldsymbol{\beta}^{*}\right):=\left\|\widehat{\boldsymbol{\beta}}-\boldsymbol{\beta}^{*}\right\|_1$. The results are summarized in Figures \ref{fig1} and \ref{fig2}, solid lines for the proposed FAQR, dashed lines with triangles for QR, and dotted lines with diamonds for FARM.
Since QR applies Lasso directly on $(\boldsymbol{X}, \boldsymbol{Y})$, it leads to much worse results, partly  due to the inadequacy of the model. Compared to FARM, FAQR performs competitively when the noise is Gaussian, but better when the noise is heavy-tailed, especially when the second moment doesn't exist ($t_2$ distribution).

To further evaluate the performance across different methods, we report the true and false positive rates (TPR and FPR), defined as the proportion of correctly estimated nonzeros and falsely estimated nonzeros, respectively. Table \ref{merged_table} summarizes the results. The performance of FAQR is competitive to FARM when the noise is Gaussian but much better when the noise is $t_2$. Compared to QR, FAQR is always better.

%\yanlin{Combine two tables: (i) keep two digits for the results, even if it is 1 (change it to 1.00); (ii) three short lines for $n=200$, $n=500$ and $n=1000$; (iii) additional column, to specify the noise distribution and dimension.}

\begin{table}[!h]%\renewcommand{\arraystretch}{0.75}
\begin{center} \tabcolsep 1.5mm\caption{TPR and FPR comparisons.}   
\begin{tabular}{cccccccc}
    \toprule
    Setting & Method & \multicolumn{2}{c}{$n=200$} & \multicolumn{2}{c}{$n=500$} & \multicolumn{2}{c}{$n=1000$} \\
    \cmidrule(lr{0.4em}){3-4} \cmidrule(lr{0.4em}){5-6} \cmidrule(lr{0.4em}){7-8}
    & & TPR & FPR & TPR & FPR & TPR & FPR \\
    \hline
    Gaussian & FAQR & 1.00(0.00) & 0.00(0.00) & 1.00(0.00) & 0.00(0.00) & 1.00(0.00) & 0.00(0.00) \\
    ($d=200$) & QR  & 0.95(0.01) & 0.08(0.00) & 1.00(0.00) & 0.10(0.00) & 1.00(0.00) & 0.07(0.00) \\
    & FARM & 1.00(0.00) & 0.03(0.00) & 1.00(0.00) & 0.03(0.00) & 1.00(0.00) & 0.03(0.00) \\
    \hline
    $t_2$ & FAQR & 1.00(0.00) & 0.00(0.00) & 1.00(0.00) & 0.00(0.00) & 1.00(0.00) & 0.00(0.00) \\
    ($d=500$) & QR  & 0.92(0.01) & 0.02(0.00) & 1.00(0.00) & 0.05(0.00) & 1.00(0.00) & 0.10(0.00) \\
    & FARM & 0.97(0.01) & 0.04(0.00) & 0.98(0.01) & 0.03(0.00) & 1.00(0.00) & 0.03(0.00) \\
       \bottomrule
    \label{merged_table}
\end{tabular}
\end{center}
\vspace{-5pt}
\footnotesize{FAQR: the proposed method; QR: convolution-smoothed QR, regressing $Y$ on $\boldsymbol{X}$ directly; FARM: method in \cite{fan2023latent}.}
\end{table}

\subsection{Adequacy of Factor Regression}
We conduct some experiments to test the hypothesis
\begin{align}
    \label{eq23}
    H_0: \boldsymbol{\beta}^{*}=0 \text { versus } H_1: \boldsymbol{\beta}^{*} \neq 0,
\end{align}
and illustrate the necessity of FAQR.

\textbf{Data Generation Processes}. We choose $n=200, M=2$ and $d$ either 200 or 500.  With the matrix $\boldsymbol{X}=\boldsymbol{F} \boldsymbol{B}^{\top}+\boldsymbol{U}$, the generating processes of $\boldsymbol{F},\boldsymbol{U},\boldsymbol{B}$ are the same as those in Section 5.1.

The response vector follows $\boldsymbol{Y}=\boldsymbol{F} \boldsymbol{\gamma}^{*}+\boldsymbol{U} \boldsymbol{\beta}^{*}+\mathcal{E}$ with each entry of $\mathcal{E} \in \mathbb{R}^n$ being generated independently from: (i) normal distribution $N\left(0,0.5^2\right)$ and (ii) $t_2$ distribution. We set $\boldsymbol{\gamma}^{*}=(0.5,0.5)^\top$ and $\boldsymbol{\beta}^{*}=(w, w, w, 0, \cdots, 0)^\top$. When $w=0$, the null hypothesis $\boldsymbol{Y}=\boldsymbol{F} \boldsymbol{\gamma}^{*}+\mathcal{E}$ holds. 

\begin{figure}
    \centering
    \includegraphics[width=1\linewidth]{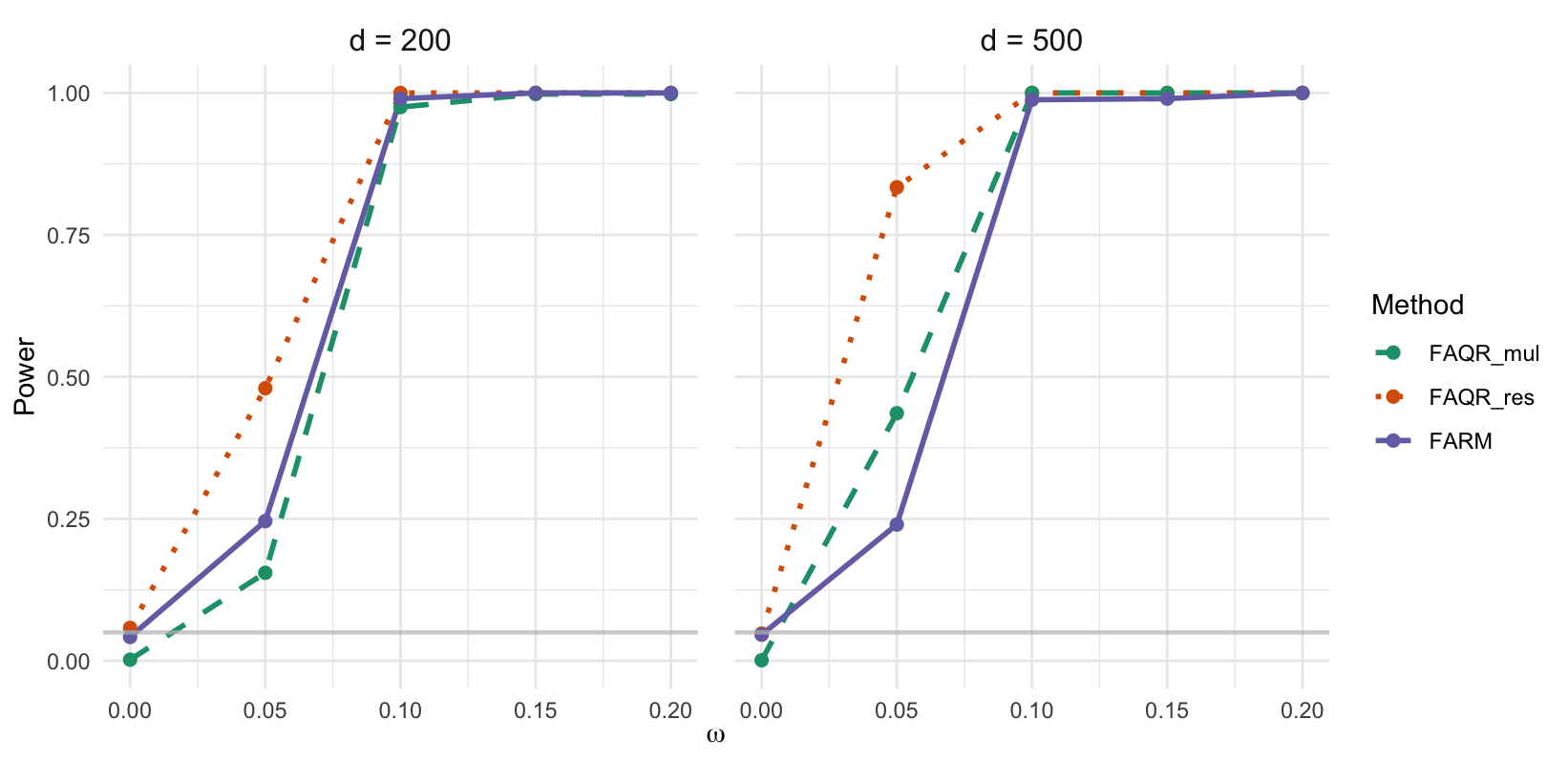}
    \caption{{Power curves of different methods with $d=200$ and $d=500$. The gray horizontal line indicates the nominal level of 0.05.}}
    \label{gau}
\end{figure}

\begin{figure}
    \centering
    \includegraphics[width=1\linewidth]{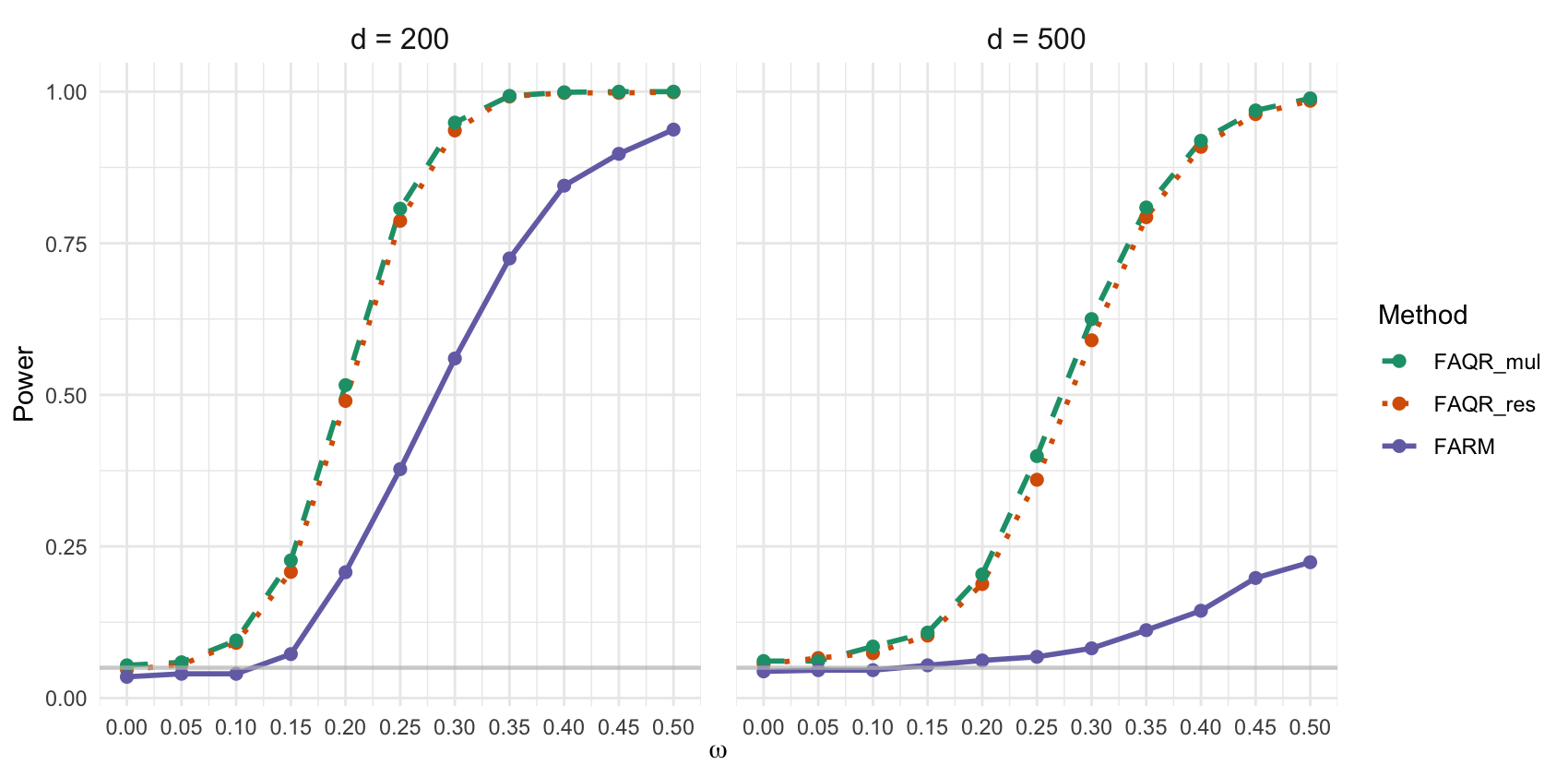}
    \caption{Power curves of different methods with $d=200$ and $d=500$. The gray horizontal line indicates the nominal level of 0.05.}
    \label{t2}
\end{figure}

Table \ref{tab:type1_error} reveals that our test gives approximately the right size. Figures \ref{gau} and \ref{t2} present the power curves of three testing procedures, FAQR with multiplier bootstrap (FAQR\_mul), FAQR with residual bootstrap (FAQR\_res) and FARM. Figure \ref{gau} shows that, under homoscedastic Gaussian errors, all three methods are comparable. However, when the noise is heavy-tailed $t_2$ distribution, the propose FAQR\_mul and FAQR\_res are much powerful than FARM.

\begin{table}[htbp]
\begin{center}
  \caption{Type I error rates under the null hypothesis.}
  \label{tab:type1_error}
  \begin{tabular}{llccc}
    \toprule
    Dimension ($d$) & Noise & \multicolumn{3}{c}{Method} \\
    \cmidrule(lr){3-5}
    & & FAQR\_mul & FAQR\_res & FARM \\
    \midrule
    \multirow{2}{*}{200} & Gaussian & - & 0.058 & 0.046 \\
                         & $t_2$    & 0.056 & 0.061 & 0.044 \\
    \multirow{2}{*}{500} & Gaussian & - & 0.048 &  0.042\\
                         & $t_2$    & 0.047 & 0.054 & 0.039 \\
    \bottomrule
  \end{tabular}
  \end{center}
  \vspace{5pt}
\footnotesize{FAQR\_mul and FAQR\_res: the proposed FAQR with multiplier bootstrap and residual bootstrap, respectively; FARM: method in \cite{fan2023latent}.}
\end{table}

\section{Real data analysis}\label{real}

\begin{figure}
    \centering
    \includegraphics[width=1\linewidth]{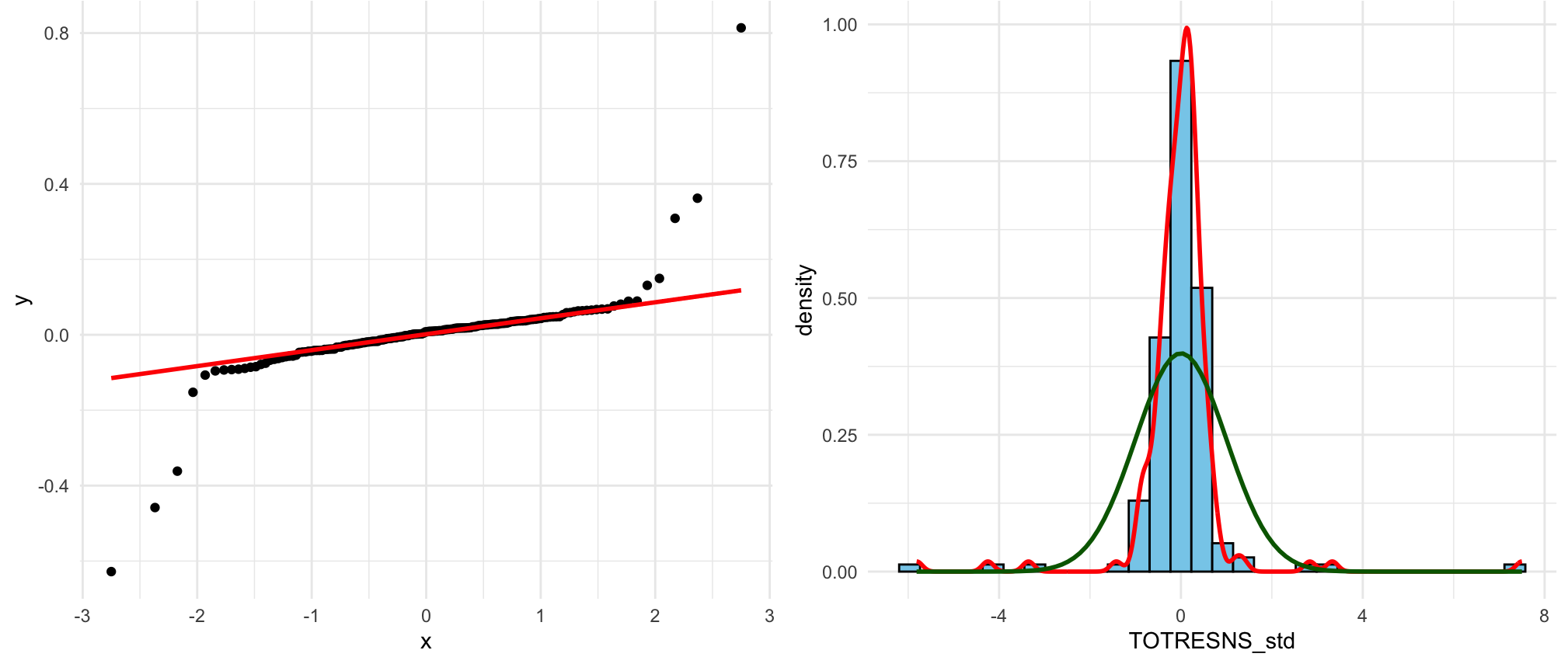}
    \caption{Q-Q plot (left) and Empirical distribution (right) of TOTRESNS (1995:1–2008:4).}
    \label{qqz}
\end{figure}

In this section, we apply the proposed method to a macroeconomic dataset FRED-MD \citep{mccracken2016fred} that includes monthly US macroeconomic indicators. Following McCracken and Ng \cite{mccracken2016fred}, each series is transformed based on its transformation code (tcode) to induce stationarity. Also, missing observations are treated using the method of ``Tall-Wide". After removing some missing values, we analyze the data from January 1997 to December 2008.
We choose ``TOTRESNS" (total reserves of depository institutions) as our response and let the remaining variables be the covariates. Figure \ref{qqz} displays the empirical distribution of TOTRESNS. The histogram and Q–Q plot (against a Gaussian) reveal that TOTRESNS is leptokurtic, exhibiting frequent extreme values. In particular, its tails are much heavier than a Gaussian distribution, which indicates ``heavy-tail".

We apply four methods to analyze the data set: (i) the proposed FAQR; (ii) FARM in Fan et al.\cite{fan2023latent}; (iii) QR, $\ell_1$ regularized convolution-smoothed QR, regressing $Y$ on the original 125 covariates directly; (iv) QR\_Factor, PCA for factor analysis, followed by quantile regression on common factors. We focus on prediction accuracy, using the rolling window method with window size 90 months, that is, using data from months 1 to 90 to predict month 91, months 2 to 91 to predict month 92, and so on. Since the data range from January 1997 to December 2008, the first prediction is at July 2004.

Model performance was evaluated by MAPE (mean of absolute prediction error) and the out-of-sample quantile pseudo-$R^2$ Koenker and Machado \citep{koenker1999goodness} 
$$R^2(\tau)=1-
\frac{\sum_i \rho_{\tau}(Y_i - \hat Y_i)}{\sum_i \rho_{\tau}(Y_i - \tilde{q}_\tau)},$$ 
where $\hat Y_i$'s are the predicted values and $
\tilde{q}_\tau$ is the sample $\tau$-quantile of $Y$. Table \ref{r2} reports the MAPE and out-of-sample pseudo-$R^2$ at $\tau=0.5$. FAQR achieved the highest pseudo-$R^2$, reflecting its superior robustness to the heavy-tailed response. All quantile-based methods lead to a smaller MAPE than FARM, partly due to the economic crisis in early 2008, which resulted in anomalies in the data, making FARM ineffective during this period.

\begin{table}[htbp]\caption{Out-of-sample $R^2$ and MAPE.}
\begin{center}
  \begin{tabular}{lcccc}
    \toprule
    Method &  FARM & FAQR &QR& QR\_Factor \\
     $R^2$ & - &0.925 &0.922  & 0.895\\  
     MAPE & 0.229 & 0.062 & 0.064 & 0.072\\
    \bottomrule
  \end{tabular}
    \label{r2}
\end{center}
\vspace{5pt}
\footnotesize{FARM: method in \cite{fan2023latent}; FAQR: the proposed method; QR: convolution-smoothed QR, regressing $Y$ on $\boldsymbol{X}$ directly; QR\_Factor: quantile regression on common factors.}
\end{table}

\begin{figure}
    \centering
    \includegraphics[width=1\linewidth]{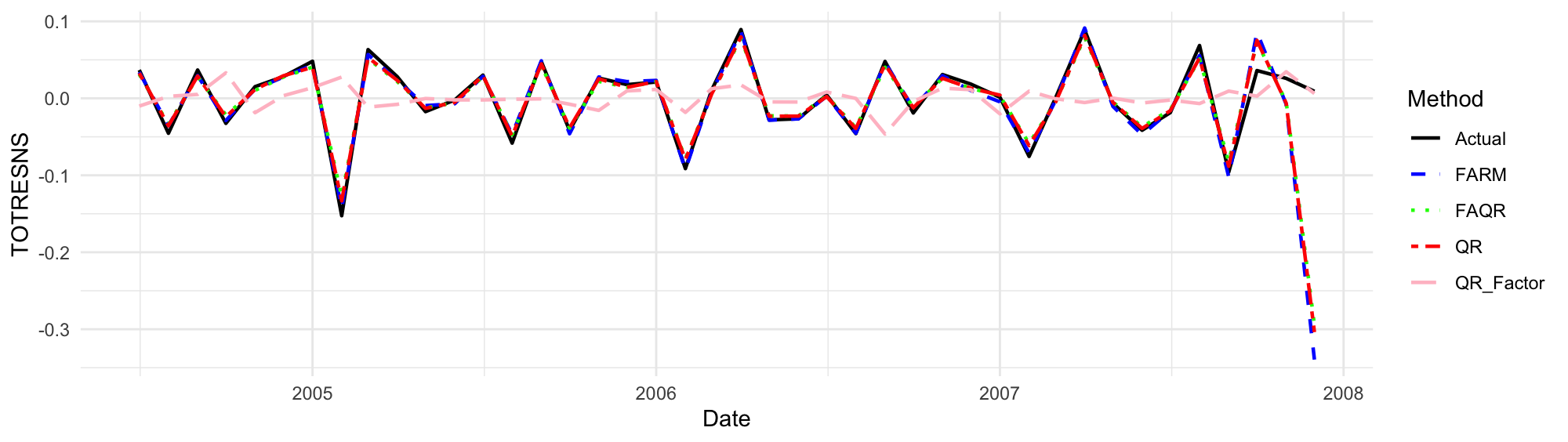}
    \caption{Actual and predicted TOTRESNS (7/2004–11/2007).}
    \label{T+1}
\end{figure}

\begin{figure}
    \centering
    \includegraphics[width=1\linewidth]{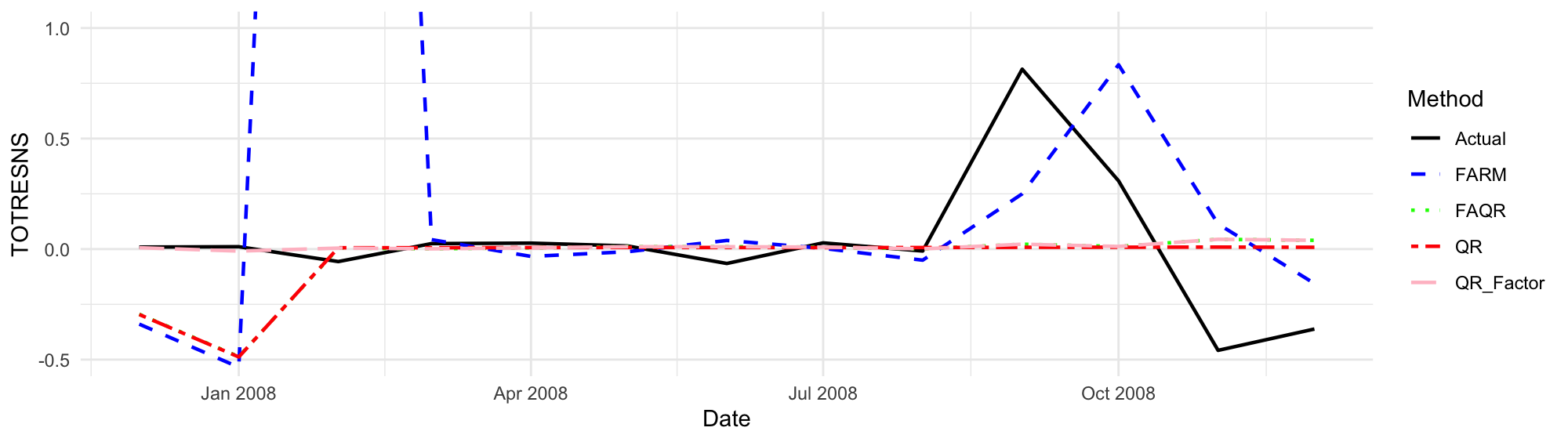}
    \caption{Actual and predicted TOTRESNS (12/2007–12/2008). The predicted value of FARM at February 2008 is too abnormal and exceeded the range of the numerical axis.}
    \label{08}
\end{figure}

We further show the prediction performance through figures. Due to the economic crisis in early 2008, we show the prediction for the period July 2004 to November 2007 and December 2007 to December 2008, separately, in Figures \ref{T+1} and \ref{08}, where the figure includes the true values and the predicted values from different methods. Figure \ref{T+1} shows that although QR\_Factor appears to perform well in MAPE and out-of-sample $R^2$, it only fits a smooth curve and does not capture the real trend. Figure \ref{08} shows that FAQR tracks the central tendency of TOTRESNS more closely than other methods, especially around the two notable outliers in the sample. In contrast, FARM is pulled toward outliers, whereas QR\_Factor and QR exhibit somewhat flatter fits.

We also use FAQR as an alternative model to conduct hypothesis tests on the adequacy of latent factor regression. The resulted $p$-values are 0.007 and 0.049 for FAQR\_res and FAQR\_mul, respectively, showing that the latent factor regression is inadequate.

In summary, this empirical analysis shows that FAQR has superior predictive performance in modeling a heavy-tailed macroeconomic variable, TOTRESNS, especially during periods of economic turmoil, resulting in higher pseudo $R^2$, lower MAPE, more accurate tracking of the actual data trajectory. Although standard quantile regression methods (QR, QR\_Factor) are moderately robust, they perform worse than FAQR in capturing the central tendency of the distribution under extreme stress, indicating the need for factor augmentation.

\section{Discussion}

This paper proposed a convolution-smoothed FAQR framework to simultaneously capture the sparse and dense effects of the features, which are possibly highly correlated. Furthermore, FAQR provides a complete picture of the heterogeneous relationships between the response and features, allowing for heavy-tailed noises.
Several promising directions for future research emerge from this work. For example, extending the factor-enhanced quantile regression analysis to survival data containing censored or truncated observations, remains a critical yet challenging issue. Moreover, while our framework primarily targets homogeneous parameters, it can be naturally extended to accommodate subgroup analysis within the factor-enhanced quantile regression setting.

\section*{Supplementary Materials}
The online Supplementary Material includes technique proofs of Theorems 3.1 and 4.1.

\section*{Acknowledgements}
Tang's work was partially supported by the National Natural Science Foundation of China, grant 12371265, and the Natural Science Foundation of Shanghai, grant 24ZR1455200; Guo's work was supported by the Fundamental Research Funds for the Central Universities; Hao's research was partially supported by Fundamental Research Funds for the Central Universities in UIBE (CXTD14-05), and the National Natural Science Foundation of China, grants 12371264 and 12171329.

\section*{Declarations}

The authors declared no potential conflicts of interest with respect to the research, authorship, and/or publication of this article.

\bibliography{sn-bibliography}% common bib file

\end{document}

\title[Factor Augmented Quantile Regression Model]{Factor Augmented Quantile Regression Model}


\author[1]{\fnm{Xiaoyang} \sur{Wei}}

\author*[1]{\fnm{Yanlin} \sur{Tang}}\email{yltang@fem.ecnu.edu.cn}

\author[2]{\fnm{Xu} \sur{Guo}}

\author[3]{\fnm{Meiling} \sur{Hao}}

\author[2]{\fnm{Yanmei} \sur{Shi}}
\affil[1]{KLATASDS-MOE, School of Statistics, East China Normal University, Shanghai, China}

\affil[2]{School of Statistics, Beijing Normal University, Beijing, China}

\affil[3]{School of Statistics, University of International Business and Economics, Beijing, China}




\maketitle

	\pagestyle{plain} \pagenumbering{arabic}
	
	\maketitle
	
	\begin{center}
		{\large \bf Supplementary Materials}
	\end{center}

	\vspace{0.1in}
	
	\def\theequation{S.\arabic{equation}}
	\def\thesection{S\arabic{section}}
	\renewcommand{\thetable}{S.\arabic{table}}
	
Recall that
$$
\begin{aligned} 
    \boldsymbol{X}_i&=\boldsymbol{B} \boldsymbol{f}_i+\boldsymbol{u}_i \notag\\ 
Q_\tau\left(Y_{i} \mid \boldsymbol{f}_i, \boldsymbol{u}_i\right)&=\boldsymbol{f}_i^{\top} \boldsymbol{\gamma}^{*}(\tau)+\boldsymbol{u}_i^{\top} \boldsymbol{\beta}^{*}(\tau), \quad i=1, \ldots, n.
\end{aligned}
$$
For notation ease, let $\boldsymbol{\theta} = (\boldsymbol{\beta}^{\top}, \boldsymbol{\gamma}^{\top})^{\top}$ and $\boldsymbol{Z} = (\boldsymbol{u}^{\top}, \boldsymbol{f}^{\top})^{\top}$.
We also define 
$$
\begin{aligned}
\nabla Q_h(\boldsymbol{\theta}^*)&= E\{\bar{K}(-\varepsilon / h)-\tau\} \boldsymbol{Z},\\
\widetilde{Q}_h(\boldsymbol{\theta})&=\frac{1}{n h} \sum_{i=1}^n \int_{-\infty}^{\infty} \rho_\tau(t) K\left(\frac{t+\boldsymbol{Z}_{i}^{\top} \boldsymbol{\theta}-Y_i}{h}\right) \mathrm{d} t,\\
\widehat{Q}_h(\boldsymbol{\theta})&=\frac{1}{n h} \sum_{i=1}^n \int_{-\infty}^{\infty} \rho_\tau(t) K\left(\frac{t+\widehat{\boldsymbol{Z}}_{i}^{\top} \boldsymbol{\theta}-Y_i}{h}\right) \mathrm{d} t.
\end{aligned}
$$
We know that $\widehat{Q}_h(\boldsymbol{\theta})$ with $\tau$-quantile is twice continuously differentiable with gradient and Hessian matrix as
$$
\begin{aligned}
\nabla_{\boldsymbol{\theta}} \widehat{Q}_h(\boldsymbol{\theta})= & \frac{1}{n} \sum_{i=1}^n\left[ \bar{K}_h(-r_i(\boldsymbol{\theta}))  -\tau \right]\widehat{\boldsymbol{Z}}_{i}, \\
 \nabla_{\boldsymbol{\theta}}^{2} \widehat{Q}_h(\boldsymbol{\theta})=&\frac{1}{n} \sum_{i=1}^n\left[K_h\left(-r_i(\boldsymbol{\theta})\right) \widehat{\boldsymbol{Z}}_{i}\widehat{\boldsymbol{Z}}_{i}^{\top}\right].
\end{aligned}
$$

\section{Proof of Theorem 3.1}

Let $\boldsymbol{H}=n^{-1} \boldsymbol{V}^{-1} \widehat{\boldsymbol{F}}^{\top} \boldsymbol{F} \boldsymbol{B}^{\top} \boldsymbol{B}$, where $\boldsymbol{V} \in \mathbb{R}^{K \times K}$ is a diagonal matrix consisting of the first $M$ largest eigenvalues of the matrix $n^{-1} \mathbb{X} \mathbb{X}^{\top}$. We first provide some useful lemmas.

\begin{lemma}\label{l1}
Under Assumptions 1-3, we have the following results.
\begin{enumerate}[i.]
        \item $\left\|\widehat{\boldsymbol{F}}-\boldsymbol{F} \boldsymbol{H}^{\top}\right\|_{\mathbb{F}}^2=O_p\left(\frac{n}{d}+\frac{1}{n}\right)$.
        \item  For any $\mathcal{I} \subseteq\{1,2, \ldots, d\}$, we have $\max _{j \in \mathcal{I}} \sum_{i=1}^n\mid \widehat{u}_{i j}-u_{i j}\mid ^2=O_p(\log \mid \mathcal{I}\mid +n / d)$.
        \item  $\left\|\boldsymbol{H}^{\top} \boldsymbol{H}-\boldsymbol{I}_M\right\|_{\mathbb{F}}^2=O_p\left(\frac{1}{n}+\frac{1}{d}\right)$.
        \item  $\max _{l \in[d]}\left\|\widehat{\boldsymbol{b}}_l-\boldsymbol{H} \boldsymbol{b}_l\right\|_2^2=O_p\left(\frac{\log d}{n}\right)$.
        \item  $\max _{i \in[n]}\left\|\widehat{\boldsymbol{f}}_i-\boldsymbol{f}_i\right\|_2=O_p\left\{\sqrt{\frac{\log n}{d}}+\sqrt{\frac{(\log d)(\log n)}{n}}\right\}$.
        \item $\frac{1}{n}\sum_{i=1}^n \left\|\widehat{\boldsymbol{f}}_{i}-\boldsymbol{f}_{i}\right\|_{2}^{2}=O_p\left(\frac{\log d}{n}+\frac{1}{d}\right).$
        \item  $\max _{i \in[n]}\left\|\widehat{\boldsymbol{u}}_i-\boldsymbol{u}_i\right\|_{\infty}=O_p\left\{\sqrt{\frac{\log n}{d}}+\sqrt{\frac{(\log d)(\log n)}{n}}\right\}$.
        \item  $\|\widehat{\boldsymbol{B}}-\boldsymbol{B}\|_{\max }=O_p\left\{\sqrt{\frac{\log d}{n}}+\frac{1}{\sqrt{d}}\right\}$.
    \end{enumerate}
\end{lemma}

Lemma \ref{l1} i.-iv. follows from Proposition 2.1 in \cite{fan2023latent}. Lemma \ref{l1} v.-viii follows from Theorem 10.4 and Corollary 10.1 in \cite{fan2020statistical}.

By Lemma \ref{l1}, $\max _{i \in[n]}\left\|\widehat{\boldsymbol{f}}_i-\boldsymbol{f}_i\right\|_2=O_p\left\{\sqrt{\frac{\log n}{d}}+\sqrt{\frac{(\log d)(\log n)}{n}}\right\}$ and $\max _{i \in[n]}\left\|\widehat{\boldsymbol{u}}_i-\boldsymbol{u}_i\right\|_{\infty}=O_p\left\{\sqrt{\frac{\log n}{d}}+\sqrt{\frac{(\log d)(\log n)}{n}}\right\}$, we have 
\begin{align}
    r&=\max _{i \in[n]}\left\|\widehat{\boldsymbol{Z}}_i-\boldsymbol{Z}_i\right\|_{\infty}=\max_{i\in [n]}\left\{\max _{i \in[n]}\left\|\widehat{\boldsymbol{u}}_i-\boldsymbol{u}_i\right\|_{\infty},\max _{i \in[n]}\left\|\widehat{\boldsymbol{f}}_i-\boldsymbol{f}_i\right\|_{\infty}\right\}
    \notag\\
    &=O_p\left\{\sqrt{\frac{\log n}{d}}+\sqrt{\frac{(\log d)(\log n)}{n}}\right\}.
\end{align}
Together with $\max _{j \in \mathcal{I}} \sum_{i=1}^n\mid \widehat{u}_{ij}-u_{ij}\mid^2=O_p(\log \mid\mathcal{I}\mid+n / d)$, we have $\max_{j\le d}\frac{1}{n}\sum_{i=1}^{n}(Z_{ij}-\widehat{Z}_{ij})^{2}=O_p\left(\frac{\log d}{n}+\frac{1}{d}\right)$.

\begin{lemma} \label{lemma A.1} Assume that Assumptions 3 and 4 hold, then
    $$
\left\|\boldsymbol{\Sigma}^{-1 / 2} \nabla Q_h(\boldsymbol{\theta}^*)\right\|_2\leq \frac{L_0}{2} \zeta_2 h^2,
$$
where $L_{0}$ is the constant of the Lipschitz condition and $\zeta_2=\int_{-\infty}^{\infty}|t|^2 K(t) \mathrm{d} t$.
\end{lemma}

Proof: $\nabla Q_h(\boldsymbol{\theta}^*)= E\{\bar{K}(-\varepsilon / h)-\tau\} \boldsymbol{Z}$. Given $\boldsymbol{Z}$, the conditional mean $\mathbb{E}\{\bar{K}(-\varepsilon / h) \mid  \boldsymbol{Z}\}$ satisfies 
$$
\begin{aligned}
\mathbb{E}\{\bar{K}(-\varepsilon / h) \mid  \boldsymbol{Z}\} & =\int_{-\infty}^{\infty} \bar{K}(-t / h) \mathrm{d} G(t) \\
& =-\frac{1}{h} \int_{-\infty}^{\infty} K(-t / h) G(t) \mathrm{d} t=\int_{-\infty}^{\infty} K(s) G(-h s) \mathrm{d} s \\
& =\tau+\int_{-\infty}^{\infty} K(s) \int_0^{-h s}\left\{g_{\varepsilon \mid \boldsymbol{Z}}(t)-g_{\varepsilon \mid \boldsymbol{Z}}(0)\right\} \mathrm{d} t \mathrm{~d} s,
\end{aligned}
$$
where $g_{\varepsilon \mid \boldsymbol{Z}}(\cdot)$ is the conditional density function. The first equation is due to integration by parts, the second equation uses variable substitution, and the third equation applies properties of the kernel density function $\int_{-\infty}^{+\infty} uK(u)du=0$.

According to Assumption 4b, $\mid g_{\varepsilon \mid \boldsymbol{Z}}(t_1)-g_{\varepsilon \mid \boldsymbol{Z}}(t_2)\mid  \leq L_0\mid t_1-t_2\mid $ for all $t_1, t_2 \in \mathbb{R}$. It follows that 
$$
\begin{aligned}
    \mid\mathbb{E} \bar{K}(-\varepsilon / h)-\tau\mid \leq \mathbb{E}\mid\mathbb{E}\{\bar{K}(-\varepsilon / h) \mid  \boldsymbol{Z}\}-\tau\mid\leq \frac{L_0}{2} \zeta_2 h^2,
\end{aligned}
$$ 
where $\zeta_k=\int_{-\infty}^{\infty}|t|^k K(t) \mathrm{d} t$ for $k \geq 1$. Consequently,
$$
\left\|\boldsymbol{\Sigma}^{-1 / 2} \nabla Q_h(\boldsymbol{\theta}^*)\right\|_2 \leq \frac{L_0}{2} \zeta_2 h^2 .
$$

\begin{lemma} \label{lemma A.2}
    (\cite{tan2022high} Lemma C.2) Assume that Assumptions 1-3 hold. Then,
$$
\|\nabla_{ \boldsymbol{\theta}} \widehat{Q}_h( \boldsymbol{\theta}^*)-\nabla_{ \boldsymbol{\theta}} \widetilde{Q}_h( \boldsymbol{\theta}^*)\|_{\infty}\ = O_p\left(\frac{\log d}{n}\right).
$$
\end{lemma}

\begin{lemma} \label{lemma A.3}Assume that Assumptions 1-3 hold. We have
    $$
\|\nabla_{\boldsymbol{\theta}} \widehat{Q}_h(\boldsymbol{\theta}^*)-\nabla_{\boldsymbol{\theta}} \widetilde{Q}_h(\boldsymbol{\theta}^*)\|_{\infty}= O_p\left(\sqrt{\frac{\log d}{n}}+\sqrt{\frac{1}{d}}+s\left(\sqrt{\frac{\log n}{d}}+\sqrt{\frac{(\log d)(\log n)}{n}}\right)\right).
$$
\end{lemma}
Proof:
$\|\nabla_{ \boldsymbol{\theta}} \widehat{Q}_h( \boldsymbol{\theta}^*)-\nabla_{ \boldsymbol{\theta}} \widetilde{Q}_h( \boldsymbol{\theta}^*)\|_{\infty}$ can further be written as
$$
\begin{aligned}
    &\|\nabla_{ \boldsymbol{\theta}} \widehat{Q}_h( \boldsymbol{\theta}^*)-\nabla_{ \boldsymbol{\theta}} \widetilde{Q}_h( \boldsymbol{\theta}^*)\|_{\infty}\\
    =&\|\frac{1}{n} \sum_{i=1}^n\left\{\bar{K}_{h}(-\hat{r}_i( \boldsymbol{\theta}^*))-\tau\right\} \hat{ \boldsymbol{Z}}_{i}-\frac{1}{n} \sum_{i=1}^n\left\{\bar{K}_{h}(-r_i(\boldsymbol{\theta}^* ))-\tau\right\}   \boldsymbol{Z}_{i}\|_{\infty}\\
= & \|\frac{1}{n} \sum_{i=1}^n\left\{(\bar{K}_{h}(-\hat{r}_i) \hat{ \boldsymbol{Z}}_{i}-\bar{K}_{h}(-r_i)   \boldsymbol{Z}_{i})-\tau(\hat{ \boldsymbol{Z}}_{i}-  \boldsymbol{Z}_{i})\right\} \|_{\infty}\quad(\text{ $r_i$ is short for $r_{i}(\boldsymbol{\theta})$)}\\
 =&\|\frac{1}{n} \sum_{i=1}^n\left\{\bar{K}_{h}(-\hat{r}_i) \hat{ \boldsymbol{Z}}_{i}-\bar{K}_{h}(-\hat{r}_i)   \boldsymbol{Z}_{i}+\bar{K}_{h}(-\hat{r}_i)   \boldsymbol{Z}_{i}-\bar{K}_{h}(-r_i)   \boldsymbol{Z}_{i}-\tau(\hat{ \boldsymbol{Z}}_{i}-  \boldsymbol{Z}_{i})\right\} \|_{\infty}\\
 \leq&\|\frac{1}{n} \sum_{i=1}^n  \boldsymbol{Z}_{i} (\bar{K}_{h}(-\hat{r}_i)-\bar{K}_{h}(-r_i) )\|_{\infty}+\|\frac{1}{n} \sum_{i=1}^n\{(\bar{K}_{h}(-\hat{r}_i)-\tau)(\hat{ \boldsymbol{Z}}_{i}-  \boldsymbol{Z}_{i})\}\|_{\infty}\\
 =&A_{1}+A_{2}
\end{aligned}
$$

\textbf{We first consider $A_{2}$.} Note that
\begin{align} \label{equ15}
    A_{2}&=\|\frac{1}{n} \sum_{i=1}^n\{(\bar{K}_{h}(-\hat{r}_i)-\tau)(\hat{ \boldsymbol{Z}}_{i}- \boldsymbol{Z}_{i})\}\|_{\infty}\notag\\
    &=\max _{j \leqslant d}\mid\frac{1}{n} \sum_{i=1}^n\{(\bar{K}_{h}(-\hat{r}_i)-\tau)(\hat{Z}_{ij}-Z_{ij})\}\mid\notag\\
    &\leq \max _{j \leqslant d}\sqrt{\frac{1}{n}\sum_{i=1}^n\{(\bar{K}_{h}(-\hat{r}_i)-\tau)^{2}}\sqrt{\frac{1}{n}\sum_{i=1}^n(\hat{Z}_{ij}-Z_{ij})^{2}}\notag\\
    & =O_p(\sqrt{\frac{\log d}{n}}+\frac{1}{\sqrt{d}}).
\end{align}
The second equality is due to the definition of the infinite norm, and the inequality sign comes from Cauchy-Schwarz's inequality. By Lemma \ref{l1}, $\max_{j\le d}\frac{1}{n}\sum_{i=1}^{n}(Z_{ij}-\widehat{Z}_{ij})^{2}=O_p\left(\frac{\log d}{n}+\frac{1}{d}\right)$. 

\textbf{We now consider $A_{1}$.} By Lemma \ref{l1}, we have
\begin{eqnarray}
r&=&\max_{i \in[n]}\left\|\widehat{ \boldsymbol{Z}}_i- \boldsymbol{Z}_i\right\|_{\infty}=\max\left\{\max _{i \in[n]}\left\|\widehat{\boldsymbol{u}}_i-\boldsymbol{u}_i\right\|_{\infty},\max _{i \in[n]}\left\|\widehat{\boldsymbol{f}}_i-\boldsymbol{f}_i\right\|_{\infty}\right\}\notag\\
&=&O_p\left\{\sqrt{\frac{\log n}{d}}+\sqrt{\frac{(\log d)(\log n)}{n}}\right\}.\notag
\end{eqnarray} 
Thus, we have
\begin{align} \label{equ12}
&\left\|\frac{1}{n} \sum_{i=1}^{n} \left[\bar{K}_{h}\left\{-\widehat{r}_{i}\left(\boldsymbol{\theta}^{*}\right)\right\}-\bar{K}_{h}\left\{-r_{i}\left(\boldsymbol{\theta}^{*}\right)\right\}\right]  \boldsymbol{Z}_{i}\right\|_{\max} \notag\\
& \leq \sup _{w \leq r}\left\|\frac{1}{n} \sum_{i=1}^{n} \left[\bar{K}_{h}\left\{-r_{i}\left(\boldsymbol{\theta}^{*}\right)+w \boldsymbol{\xi}_{i}^{\top} \boldsymbol{\theta}^{*}\right\}-\bar{K}_{h}\left\{-r_{i}\left(\boldsymbol{\theta}^{*}\right)\right\}\right] \boldsymbol{Z}_{i}\right\|_{\max} \notag\\
& \leq \left\|\boldsymbol{\Sigma}^{\frac{1}{2}}\right\|_{1}\left\{\underbrace{\sup _{w \leq r}\left\|\frac{1}{n} \sum_{i=1}^{n}(1-\mathbb{E}) \left[\bar{K}_{h}\left\{-r_{i}\left(\boldsymbol{\theta}^{*}\right)+w \boldsymbol{\xi}_{i}^{\top} \boldsymbol{\theta}^{*}\right\}-\bar{K}_{h}\left\{-r_{i}\left(\boldsymbol{\theta}^{*}\right)\right\}\right] \boldsymbol{W}_{i}\right\|_{\max}}_{I_{1,1}}\right. \notag\\
&\left.+\underbrace{\sup _{w \leq r}\left\|\frac{1}{n} \sum_{i=1}^{n} \mathbb{E} \left[\bar{K}_{h}\left\{-r_{i}\left(\boldsymbol{\theta}^{*}\right)+w \boldsymbol{\xi}_{i}^{\top} \boldsymbol{\theta}^{*}\right\}-\bar{K}_{h}\left\{-r_{i}\left(\boldsymbol{\theta}^{*}\right)\right\}\right] \boldsymbol{W}_{i}\right\|_{\max}}_{I_{1,2}}\right\}, 
\end{align}
where $\boldsymbol{\xi}_{i}$ is a $(d+M)$-vector with every elementary with uniform distribution in $(-1,1)$ and independent of $\boldsymbol{W}$, where $\boldsymbol{W}_i=\boldsymbol{\Sigma}^{-\frac{1}{2}}  \boldsymbol{Z}_i$. By Taylor expansion and the definition of $\|\cdot\|_{\max}$, we have

\begin{align}\label{equ13}
& \left\|\frac{1}{n} \sum_{i=1}^{n} \mathbb{E} \left[\bar{K}_{h}\left\{-r_{i}\left(\boldsymbol{\theta}^{*}\right)+w \boldsymbol{\xi}_{i}^{\top} \boldsymbol{\theta}^{*}\right\}-\bar{K}_{h}\left\{-r_{i}\left(\boldsymbol{\theta}^{*}\right)\right\}\right] \boldsymbol{Z}_{i}\right\|_{\max} \notag\\
= & \max_{1\leq j \leq d}\mid \frac{1}{n} \sum_{i=1}^{n} \mathbb{E} \left[\bar{K}_{h}\left\{-r_{i}\left(\boldsymbol{\theta}^{*}\right)+w \boldsymbol{\xi}_{i}^{\top} \boldsymbol{\theta}^{*}\right\}-\bar{K}_{h}\left\{-r_{i}\left(\boldsymbol{\theta}^{*}\right)\right\}\right]z_{ij}\mid  \notag\\
= & \max_{1\leq j \leq d}\mid \mathbb{E} \int_{-\infty}^{\infty} \left\{\bar{K}_{h}\left(\boldsymbol{Z}^{\top} \boldsymbol{\theta}^{*}+w \boldsymbol{\xi}_{i}^{\top} \boldsymbol{\theta}^{*}-y\right)-\bar{K}_{h}\left(\boldsymbol{Z}^{\top} \boldsymbol{\theta}^{*}-y\right)\right\} g_{y}\left(y \mid \boldsymbol{\xi}_{i}, \boldsymbol{Z}\right) d y z_{ij} \mid \notag\\
= &  \max_{1\leq j \leq d}\mid \mathbb{E} \int_{-\infty}^{\infty} \left\{\bar{K}\left(v+w \boldsymbol{\xi}_{i}^{\top} \boldsymbol{\theta}^{*} / h\right)-\bar{K}(v)\right\} g_{y}\left(\boldsymbol{Z}^{\top} \boldsymbol{\theta}^{*}-hv \mid \boldsymbol{\xi}_{i}, \boldsymbol{Z}\right) d yz_{ij}\mid  \notag\\
\leq & \bar{g} \max_{1\leq j \leq d}\mid \mathbb{E} \int_{-\infty}^{\infty} \int_{0}^{1} \left\{K\left(v+\varepsilon w \boldsymbol{\xi}_{i}^{\top} \boldsymbol{\theta}^{*} / h\right)\right\} d y d \varepsilon z_{ij} w \boldsymbol{\xi}_{i}^{\top} \boldsymbol{\theta}^{*}\mid  \notag\\
\lesssim & sr . 
\end{align}

To get the order of $I_{1,1}$, we define
$$
\mathcal{S}(w)=\frac{1}{n} \sum_{i=1}^{n}(1-\mathbb{E})  \bar{K}_{h}\left\{-r_{i}\left(\boldsymbol{\theta}^{*}\right)+w \boldsymbol{\xi}_{i}^{\top} \boldsymbol{\theta}^{*}\right\} \boldsymbol{W}_{i}, w \in \mathbb{R},
$$
and $\Psi(w)=\mathcal{S}(w)-\mathcal{S}(0)$. It is easy to show that $\Psi(0)=\mathbf{0}$, $\mathbb{E} \Psi(w)=\mathbf{0}$, and
$$
\begin{aligned}
\sup _{w \leq r}\left\|\frac{1}{n} \sum_{i=1}^{n}(1-\mathbb{E}) \left[\bar{K}_{h}\left\{-r_{i}\left(\boldsymbol{\theta}^{*}\right)+w \boldsymbol{\xi}_{i}^{\top} \boldsymbol{\theta}^{*}\right\}-\bar{K}_{h}\left\{-r_{i}\left(\boldsymbol{\theta}^{*}\right)\right\}\right] \boldsymbol{W}_{i}\right\|_{\max}=\sup _{w \leq r}\|\Psi(w)\|_{2}.
\end{aligned}
$$

Simple calculations yields
$$
\begin{aligned}
\nabla \Psi(w) & =\frac{1}{n} \sum_{i=1}^{n}(1-\mathbb{E})  K_{h}\left\{-r_{i}\left(\boldsymbol{\theta}^{*}\right)+w \boldsymbol{\xi}_{i}^{\top} \boldsymbol{\theta}^{*}\right\} \boldsymbol{\xi}_{i}^{\top} \boldsymbol{\theta}^{*} \boldsymbol{z}_{i} \\
& \equiv \frac{1}{n} \sum_{i=1}^{n}(1-\mathbb{E}) \left\{\phi_{i}(w) \boldsymbol{\xi}_{i}^{\top} \boldsymbol{\theta}^{*} \boldsymbol{z}_{i}\right\},
\end{aligned}
$$
where $\phi_{i}(w)= K_{h}\left\{-r_{i}\left(\boldsymbol{\theta}^{*}\right)+w \boldsymbol{\xi}_{i}^{\top} \boldsymbol{\theta}^{*}\right\}$. It is easy to verify $0 \leq \phi_{i}(w) \leq \kappa / h$ with $\kappa_{u}=\|K\|_{\infty}$. For any $\mathbf{u} \in \mathbb{S}^{d+M-1}$ and $|v| \leq \min \left(n h /\left(\kappa_{u}\left\|\boldsymbol{\theta}^{*}\right\|_{\infty}\right), n / \bar{g}\right)$, by independence and the elementary inequality $e^{x} \leq 1+x+x^{2} e^{|x|} / 2$, we obtain that
\begin{align}
& \mathbb{E} \exp \left\{v \boldsymbol{\alpha}^{\top} \nabla \Psi(w)\right\} \notag\\
 \leq&\left[1+\frac{v^{2}}{2 n^{2}} e^{\frac{\bar{g}|v|}{n}} \mathbb{E}\mid \boldsymbol{W}^{\top} \boldsymbol{\alpha} \boldsymbol{\xi}^{\top} \boldsymbol{\theta}^{*}\mid  \mathbb{E}\left\{\phi_{\boldsymbol{v}} \boldsymbol{W}^{\top} \boldsymbol{\alpha} \boldsymbol{\xi}^{\top} \boldsymbol{\theta}^{*}-\mathbb{E}\left(\phi_{\boldsymbol{v}} \boldsymbol{W}^{\top} \boldsymbol{\alpha} \boldsymbol{\xi}^{\top} \boldsymbol{\theta}^{*}\right)\right\}^{2}\right. \notag\\
& \left.e^{\frac{\kappa u|v|}{n h}}\mid \boldsymbol{W}^{\top} \boldsymbol{\alpha} \boldsymbol{\xi}^{\top} \boldsymbol{\theta}^{*}\mid \right]^{n} \notag\\
 \stackrel{(i)}{\lesssim}&\left[1+\frac{v^{2}}{2 n^{2}} e^{\bar{g}|v| / n} \mathbb{E}\left\{\phi_{\boldsymbol{v}} \boldsymbol{W}^{\top} \boldsymbol{\alpha} \boldsymbol{\xi}^{\top} \boldsymbol{\theta}^{*}-\mathbb{E}\left(\phi_{\boldsymbol{v}} \boldsymbol{W}^{\top} \boldsymbol{\alpha} \boldsymbol{\xi}^{\top} \boldsymbol{\theta}^{*}\right)\right\}^{2} e^{\frac{\kappa u|v|}{n h}\mid \boldsymbol{W}^{\top} \boldsymbol{\alpha} \boldsymbol{\xi}^{\top} \boldsymbol{\theta}^{*}\mid }\right]^{n} \notag\\
\leq& \left[1+\frac{e v^{2}}{2 n^{2}} \mathbb{E}\left\{\phi_{\boldsymbol{v}} \boldsymbol{W}^{\top} \boldsymbol{\alpha} \boldsymbol{\xi}^{\top} \boldsymbol{\theta}^{*}-\mathbb{E}\left(\phi_{\boldsymbol{v}} \boldsymbol{W}^{\top} \boldsymbol{\alpha} \boldsymbol{\xi}^{\top} \boldsymbol{\theta}^{*}\right)\right\}^{2} e^{\kappa u \zeta^{2}|v| /(n h)}\right]^{n} \notag\\ \label{equ11}
 \leq&\left\{1+\frac{(e v)^{2}}{2 n^{2}} \mathbb{E}\left(\phi_{\boldsymbol{v}} \boldsymbol{W}^{\top} \boldsymbol{\alpha} \boldsymbol{\xi}^{\top} \boldsymbol{\theta}^{*}\right)^{2}\right\}^{n}, 
\end{align}
where inequality (i) follows from the bound $\mathbb{E}\mid \boldsymbol{W}^{\top} \boldsymbol{\alpha} \boldsymbol{\xi}^{\top} \boldsymbol{\theta}^{*}\mid  \leq\left\|\boldsymbol{\theta}^{*}\right\|_{\infty}$. For\\ $\phi_{i}(w)= K_{h}\left\{-r_{i}\left(\boldsymbol{\theta}^{*}\right) w \boldsymbol{\xi}_{i}^{\top} \boldsymbol{\theta}^{*}\right\}$, its conditional second moment can be bounded by
$$
\begin{aligned}
\mathbb{E}\left(\phi_{i}(w)^{2} \mid \boldsymbol{W}, \boldsymbol{\xi}_{i}\right) & =\frac{1}{h^{2}} \int_{-\infty}^{\infty} K^{2}\left(\frac{\boldsymbol{W}^{\top} \boldsymbol{\theta}^{*}-t}{h}\right) g_{y}\left(t \mid \boldsymbol{W}, \boldsymbol{\xi}_{i}\right) d t \\
& =\frac{1}{h} \int_{-\infty}^{\infty} K^{2}(u) g_{y}\left(\boldsymbol{W}^{\top} \boldsymbol{\theta}^{*}+u \mid \boldsymbol{W}, \boldsymbol{\xi}_{i}\right) d u \leq \frac{\kappa_{u} \bar{g}}{h}.
\end{aligned}
$$
Substituting this into \eqref{equ11} yields
\begin{eqnarray}
\mathbb{E} \exp \left\{v \boldsymbol{\alpha}^{\top} \nabla \Psi(w)\right\} &\leq&\left\{1+\kappa_{u} \bar{g} e^{2} v^{2}\left\|\boldsymbol{\theta}^{*}\right\|_{\infty} /\left(2 n^{2} h\right)\right\}^{n} \notag\\
&\leq& \exp \left\{\kappa_{u} \bar{g}\left\|\boldsymbol{\theta}^{*}\right\|_{\infty} e^{2} v^{2} /(2 n h)\right\}.\notag
\end{eqnarray}
This verifies condition (A.4) in \cite{spokoiny2014bernsteinvonmises}. Therefore, applying Theorem A. 3 therein, we obtain that with probability at least $1-d^{-1}$,
\begin{equation} \label{equ13}
\sup _{w \leq r}\|\Psi(\boldsymbol{v})\|_{2} \lesssim\left(\kappa_{u} \bar{g}\right)^{1 / 2} \sqrt{\frac{\log d}{n h}} r 
\end{equation}
as far as $n h \gtrsim \sqrt{\log d}$. Then, combining \eqref{equ12}-\eqref{equ13}, we can get
\begin{align} \label{equ14}
& \left\|\frac{1}{n} \sum_{i=1}^{n} \left[\bar{K}_{h}\left\{-r_{i}\left(\boldsymbol{\theta}^{*}\right)\right\}-\bar{K}_{h}\left\{-r_{i}\left(\boldsymbol{\theta}^{*}\right)\right\}\right] \boldsymbol{Z}_i\right\|_{\Sigma^{-1}} \notag\\
= & O_p\left(\left(s+\sqrt{\frac{\log d}{n h}}\right)\left(\sqrt{\frac{\log n}{d}}+\sqrt{\frac{(\log d)(\log n)}{n}}\right)\right) \notag\\
= & O_p\left(s\left(\sqrt{\frac{\log n}{d}}+\sqrt{\frac{(\log d)(\log n)}{n}}\right)\right).
\end{align}

Combing \eqref{equ15} and \eqref{equ14}, we have
\begin{eqnarray}
&&\|\nabla_{\boldsymbol{\theta}} \widehat{Q}_h(\boldsymbol{\theta}^*)-\nabla_{\boldsymbol{\theta}} \widetilde{Q}_h(\boldsymbol{\theta}^*)\|_{\infty}\notag\\
&=& O_p\left(\sqrt{\frac{\log d}{n}}+\sqrt{\frac{1}{d}}+s\left(\sqrt{\frac{\log n}{d}}+\sqrt{\frac{(\log d)(\log n)}{n}}\right)\right).\notag
\end{eqnarray}

\begin{lemma}\label{lemma A.4} Assume that Assumptions 1-4 hold, we have
\begin{eqnarray}
&&\frac{1}{n} \sum_{i=1}^n K_{h}(-\widehat{r}_{i}(\boldsymbol{\theta}^{*}))\hat{\boldsymbol{Z}}_{i} \hat{\boldsymbol{Z}}_{i}^{\top}-E\frac{1}{n} \sum_{i=1}^n K_{h}(-\varepsilon_{i})\boldsymbol{Z}_{i} \boldsymbol{Z}_{i}^{\top}\notag\\
&=&O_p\left(\left(sr\left(\frac{1}{h}+2 \sqrt{ \frac{\log d}{n h^3}}+\frac{\log d}{n h^2}\right)+\frac{s^2 r^2\log (2nd^{2})}{h^3}\right)\right). \notag   
\end{eqnarray}
\end{lemma}

Proof: $\frac{1}{n} \sum_{i=1}^n K_{h}(-\widehat{r}_{i}(\boldsymbol{\theta}^{*}))\hat{\boldsymbol{Z}}_{i} \hat{\boldsymbol{Z}}_{i}^{\top}-E\frac{1}{n} \sum_{i=1}^n K_{h}(-\varepsilon_{i})\boldsymbol{Z}_{i} \boldsymbol{Z}_{i}^{\top}$ can further be written as
$$
\begin{aligned}
    &\frac{1}{n} \sum_{i=1}^n K_{h}(-\widehat{r}_{i}(\boldsymbol{\theta}^{*}))\hat{\boldsymbol{Z}}_{i} \hat{\boldsymbol{Z}}_{i}^{\top}-E\frac{1}{n} \sum_{i=1}^n K_{h}(-\varepsilon_{i})\boldsymbol{Z}_{i} \boldsymbol{Z}_{i}^{\top}\\
    =&\frac{1}{n} \sum_{i=1}^n K_{h}(-\widehat{r}_{i}(\boldsymbol{\theta}^{*}))\hat{\boldsymbol{Z}}_{i} \hat{\boldsymbol{Z}}_{i}^{\top}-\frac{1}{n} \sum_{i=1}^n K_{h}(-r_{i}(\boldsymbol{\theta}^{*}))\boldsymbol{Z}_{i} \boldsymbol{Z}_{i}^{\top}\\
    &+\frac{1}{n} \sum_{i=1}^n K_{h}(-r_{i}(\boldsymbol{\theta}^{*}))\boldsymbol{Z}_{i}\boldsymbol{Z}_{i}^{\top}-E\frac{1}{n} \sum_{i=1}^n K_{h}(-\varepsilon_{i})\boldsymbol{Z}_{i} \boldsymbol{Z}_{i}^{\top}\\
    =&A_{1}+A_{2}.
\end{aligned}
$$

\textbf{We first consider $A_2$.} Note that
$$
\begin{aligned}
A_2 & =\left\|\frac{1}{n} \sum_{i=1}^n K_h\left(-r_i\right) \boldsymbol{Z}_i \boldsymbol{Z}_i^{\top}-\mathbb{E}\left[K_h(-\varepsilon) \boldsymbol{Z} \boldsymbol{Z}^{\top}\right]\right\|_{\max} \\
& \leq\left\|\boldsymbol{\Sigma}^{\frac{1}{2}}\right\|_{1}^2 \cdot\left\|\frac{1}{n} \sum_{i=1}^n K_h\left(-r_i\right) \boldsymbol{W}_i \boldsymbol{W}_i^{\top}-\mathbb{E}\left[K_h(-\varepsilon) \boldsymbol{W}\boldsymbol{W}^{\top}\right]\right\|_{\max} \\
& :=\left\|\boldsymbol{\Sigma}^{\frac{1}{2}}\right\|_{1}^2 \cdot\left\|\frac{1}{n} \sum_{i=1}^n(1-\mathbb{E}) \phi_i \boldsymbol{W}_i \boldsymbol{W}_i^{\top}\right\|_{\max},
\end{aligned}
$$
where $\boldsymbol{W}_i=\boldsymbol{\Sigma}^{-\frac{1}{2}} \boldsymbol{Z}_i$ and $\phi_i=K_h\left(-r_i\right)$. It is easy to verify $\mid \phi_i\mid  \leq \frac{\kappa_u}{h}$ and
$$
\mathbb{E}\left(\phi_i^2 \mid \boldsymbol{Z}_i\right)=\frac{1}{h^2} \int_{-\infty}^{\infty} K^2\left(-\frac{u}{h}\right) g_{\varepsilon \mid \boldsymbol{Z}}(u) \mathrm{d} u=\frac{1}{h} \int_{-\infty}^{\infty} K^2(v) g_{\varepsilon \mid \boldsymbol{Z}}(-h v) \mathrm{d} v \leq \frac{\bar{g} \kappa_u}{h} .
$$

Denote the $j$-th element of $\boldsymbol{W}_i$ as $w_{i, j}$. Then for arbitrary pair $(j, k) \in\{1, \ldots, d\} \times\{1, \ldots, d\}$, we bound the higher order moments of $\phi_i w_{i, j} w_{i, k}$ by
\begin{align}
    &\mathbb{E}\mid \phi_i w_{i, j} w_{i, k}\mid ^m=\mathbb{E}\mid \phi_i\left(\boldsymbol{e}_j^{\top} \boldsymbol{W}_i\right)\left(\boldsymbol{e}_k^{\top} \boldsymbol{W}_i\right)\mid ^m \notag\\
    \leq &\frac{\bar{g} \kappa_u}{h} \cdot\left(\frac{\kappa_u}{h}\right)^{m-2} \cdot \mathbb{E}\mid \boldsymbol{e}_j^{\top} \boldsymbol{W}_i\mid ^m \cdot \mathbb{E}\mid \boldsymbol{e}_k^{\top} \boldsymbol{W}_i\mid ^m .\notag
\end{align}
By Lyapunov's inequality,
\begin{align}
    &\left(\mathbb{E}\mid \boldsymbol{e}_j^{\top} \boldsymbol{W}_i\mid ^m\right)^2\notag \\
    \leq & \mathbb{E}\mid \boldsymbol{e}_j^{\top} \boldsymbol{W}_i\mid ^{2 m} \leq \sigma^{2 m} \cdot 2 m \int_0^{\infty} t^{2 m-1} \mathbb{P}\left(\mid \left\langle\boldsymbol{e}_j, \boldsymbol{W}_i\right\rangle\mid  \geq \sigma t\right) \mathrm{d} t \notag \\
    \leq & 2^{m+1} \sigma^{2 m} m!.
\end{align}
Hence, the Bernstein's condition can be verified as
$$
\mathbb{E}\mid \phi_i w_{i, j} w_{i, k}\mid ^m \leq \frac{m!}{2} \cdot\left(4 \sigma^2\right)^2 \frac{\bar{g} \kappa_u}{h} \cdot\left(\frac{2 \sigma^2 \kappa_u}{h}\right)^{m-2}
$$
for $m \geq 2$. Applying Bernstein's inequality and taking union bound over all $d^2$ pairs, we obtain that for any $t>0$, the following inequality holds with probability at least $1-2 d^2 e^{-t}$ :
\begin{align}
   & \left\|\frac{1}{n} \sum_{i=1}^n(1-\mathbb{E}) \phi_i \boldsymbol{W}_i \boldsymbol{W}_i^{\top}\right\|_{\max}\notag\\
   =&\max _{1 \leq j, k \leq d}\mid \frac{1}{n} \sum_{i=1}^n(1-\mathbb{E}) \phi_i w_{i, j} w_{i, k}\mid  \notag\\
   \leq& 2 \sigma^2\left(2 \sqrt{2 \bar{g} \kappa_u \frac{t}{n h}}+\kappa_u \frac{t}{n h}\right) .
\end{align}
Let $t=\log 2+3 \log d$, then it follows that with probability at least $1-\frac{1}{d}$ we have
\begin{eqnarray}
A_2 &\leq&\left\|\boldsymbol{\Sigma}^{\frac{1}{2}}\right\|_{1}^2 \cdot\left\|\frac{1}{n} \sum_{i=1}^n(1-\mathbb{E}) \phi_i \boldsymbol{W}_i \boldsymbol{W}_i^{\top}\right\|_{\max}\notag\\
 &\leq& 8\left\|\boldsymbol{\Sigma}^{\frac{1}{2}}\right\|_{1}^2 \sigma^2\left(\sqrt{2 \bar{g} \kappa_u \frac{\log d}{n h}}+\kappa_u \frac{\log d}{n h}\right).\notag
\end{eqnarray}

\textbf{We now consider $A_1$.} We have
$$
\begin{aligned}
A_1\leq & \underbrace{\left\|\frac{1}{n} \sum_{i=1}^n  K_h\left\{-\widehat{r}_{i}\left(\boldsymbol{\theta}^{*}\right)\right\}\left(\boldsymbol{Z}_i \boldsymbol{Z}_i^{\top}-\widehat{\boldsymbol{Z}}_i \widehat{\boldsymbol{Z}}_i^{\top}\right)\right\|_{\max}}_{U_{1,1}} \\
& +\underbrace{\left\|\frac{1}{n} \sum_{i=1}^n \left[K_h\left\{-r_{i}\left(\boldsymbol{\theta}^{*}\right)\right\}-K_h\left\{-\widehat{r}_{i}\left(\boldsymbol{\theta}^{*}\right)\right\}\right] \boldsymbol{Z}_i \boldsymbol{Z}_i^{\top}\right\|_{\max}}_{U_{1,2}} .
\end{aligned}
$$
For $U_{1,1}$, we have
\begin{align} \label{equ2}
U_{1,1}=& \max _{j,k \leq d}\mid \frac{1}{n} \sum_{t=1}^n K_{h}(\widehat{r}_{i}(\boldsymbol{\theta}^{*}))\left(\hat{z}_{ij} \hat{z}_{ik}-z_{ij} z_{ik}\right)\mid  \notag\\
\leq & \max _{j,k \leq d}\mid \frac{1}{n} \sum_{t=1}^n K_{h}(\widehat{r}_{i}(\boldsymbol{\theta}^{*}))\left(\hat{z}_{ij}-z_{ij}\right)\left(\hat{z}_{ik}-z_{ik}\right)\mid \notag\\
&+2 \max _{j,k \leq d}\mid \frac{1}{n} \sum_{t=1}^n  K_{h}(\widehat{r}_{i}(\boldsymbol{\theta}^{*}))z_{ij}\left(\hat{z}_{ik}-z_{ik}\right)\mid  \notag\\
\leq & \mid K_{h}(\widehat{r}_{i}(\boldsymbol{\theta}^{*}))\mid \max _{j \leq d} \frac{1}{n} \sum_{t=1}^n\left(\hat{z}_{ij}-z_{ij}\right)^2\notag\\
&+2 \sqrt{\max _{i \leq d} \frac{1}{n} \sum_{t=1}^n z_{ij}^2} \sqrt{\mid K_{h}(\widehat{r}_{i}(\boldsymbol{\theta}^{*}))\mid ^{2}\max _{i \leq d} \frac{1}{n} \sum_{t=1}^n\left(\hat{z}_{ij}-z_{ij}\right)^2} \notag\\
= & O_p\left(\frac{\log d}{nh}+\sqrt{\frac{\log d}{nh^{2}}}\right).
\end{align}

Next, we derive the upper bound of $U_{1,2}$. We have
\begin{align} 
& U_{1,2}=\left\|\frac{1}{n} \sum_{i=1}^n \left[K_h\left\{-r_{i}\left(\boldsymbol{\theta}^{*}\right)\right\}-K_h\left\{-\widehat{r}_{i}\left(\boldsymbol{\theta}^{*}\right)\right\}\right] \boldsymbol{Z}_i \boldsymbol{Z}_i^{\top}\right\|_{\max} \notag\\
& =\mid \frac{1}{n} \sum_{i=1}^n\left[\frac{1}{h^2} K^{\prime}\left(-\frac{\varepsilon_i}{h}\right)\left((\widehat{\boldsymbol{Z}}_{i}-\boldsymbol{Z}_i)^{\top} \boldsymbol{\theta}^{*}\right)+\frac{1}{2 h^3} K^{\prime \prime}\left(\eta_i\right)\left((\widehat{\boldsymbol{Z}}_{i}-\boldsymbol{Z}_i)^{\top} \boldsymbol{\theta}^{*}\right)^2\right] z_{i, j} z_{i, k}\mid  \notag\\
& \leq\mid \frac{1}{n h^2} \sum_{i=1}^n K^{\prime}\left(-\frac{\varepsilon_i}{h}\right) z_{i, j} z_{i, k}\left((\widehat{\boldsymbol{Z}}_{i}-\boldsymbol{Z}_i)^{\top} \boldsymbol{\theta}^{*}\right)\mid \notag\\
&+\mid \frac{1}{2 n h^3} \sum_{i=1}^n K^{\prime \prime}\left(\eta_i\right) z_{i, j} z_{i, k}\left((\widehat{\boldsymbol{Z}}_{i}-\boldsymbol{Z}_i)^{\top} \boldsymbol{\theta}^{*}\right)^2\mid  \notag\\
& :=J_1+J_2.\notag
\end{align}

Under Assumption 4, $\mid K^{\prime \prime}(u)\mid  \leq \kappa_u^{\prime \prime}$ for any $u \in \mathbb{R}$. Then the upper bound of $J_2$ is
\begin{align} \label{equ24}
J_2 & =\max _{1 \leq j, k \leq d}\mid \frac{1}{2 n h^3} \sum_{i=1}^n K^{\prime \prime}\left(\eta_i\right) z_{i, j} z_{i, k}\left((\widehat{\boldsymbol{Z}}_{i}-\boldsymbol{Z}_i)^{\top} \boldsymbol{\theta}^{*}\right)^2\mid  \notag\\
& \leq \max _{1 \leq j, k \leq d} \frac{\kappa_u^{\prime \prime}}{2 h^3}\frac{1}{n} \sum_{i=1}^n \mid z_{i, j} z_{i, k}\left(\left\|\boldsymbol{\theta}^{*}\right\|_{1}\left\|\widehat{\boldsymbol{Z}}_{i}-\boldsymbol{Z}_i\right\|_{\infty}\right)^2\mid  \notag\\
& \leq \frac{\kappa_u^{\prime \prime}}{2 h^3}\max_{i\leq n}\left\|\widehat{\boldsymbol{Z}}_{i}-\boldsymbol{Z}_i\right\|_{\infty}^2 \cdot \left\|\boldsymbol{\theta}^{*}\right\|_{1}^2 \cdot \max _{1 \leq j, k \leq d}\frac{1}{n} \sum_{i=1}^n  \mid z_{i, j} z_{i, k}\mid  \notag\\
& \leq \frac{\kappa_u^{\prime \prime}}{2 h^3}\max_{i\leq n}\left\|\widehat{\boldsymbol{Z}}_{i}-\boldsymbol{Z}_i\right\|_{\infty}^2 \cdot \max _{\substack{1 \leq i \leq n \\
1 \leq j \leq d}}\mid z_{i, j}\mid ^2\cdot \left\|\boldsymbol{\theta}^{*}\right\|_{1}^2,
\end{align}
where the first inequality is derived from the Hölder's inequality. The sub-Gaussian nature of $z_{i, j}$ leads to that for every $t>0$,

$$
\mathbb{P}\left(\max _{\substack{1 \leq i \leq n \\ 1 \leq j \leq d}}\mid z_{i, j}\mid  \geq t\right) \leq \sum_{i=1}^n \sum_{j=1}^d \mathbb{P}\left(\mid z_{i, j}\mid  \geq t\right) \leq 2 n d e^{-\frac{t^2}{2 \sigma^2}}.
$$
Setting $t=\sigma \sqrt{2 \log \left(2 n d^2\right)}$, we have
\begin{align} \label{equ1}
    \max _{\substack{1 \leq i \leq n \\ 1 \leq j \leq d}}\mid z_{i, j}\mid  \leq \sigma \sqrt{2 \log \left(2 n d^2\right)}
\end{align}
with probability at least $1-\frac{1}{d}$. Recall $r$ as the $\infty$-norm error bound of $\widehat{\boldsymbol{Z}}_{i}-\boldsymbol{Z}_i$. Inserting \eqref{equ1} and $r$ into \eqref{equ24} yields that, the following upper bound of $J_2$ holds with probability at least $1-\frac{2}{d}$,
\begin{align} \label{eq8}
    \ J_2 \leq \frac{\kappa_u^{\prime \prime}}{2 h^3} \cdot r^2 \cdot \sigma^2 \cdot 2 \log \left(2 n d^2\right) \cdot s^2=O_p\left(\frac{s^2 r^2\log (2nd^{2})}{h^3}\right).
\end{align}

Finally, we derive the upper bound of $J_1$. It is easy to obtain that
\begin{align} \label{eq5}
J_1 & =\mid \frac{1}{n h^2} \sum_{i=1}^n K^{\prime}\left(-\frac{\varepsilon_i}{h}\right) z_{i, j} z_{i, k}\left(\widehat{\boldsymbol{Z}}_i-\boldsymbol{Z}_{i}\right)^{\top} \boldsymbol{\theta}^{*}\mid  \notag\\
& \leq \left\|\boldsymbol{\theta}^{*}\right\|_1 \cdot \max _{1 \leq i \leq n}\left\|\widehat{\boldsymbol{Z}}_{i}-\boldsymbol{Z}_i\right\|_{\infty} \cdot\mid \frac{1}{n h^2} \sum_{i=1}^n K^{\prime}\left(-\frac{\varepsilon_i}{h}\right) z_{i, j} z_{i, k}\mid ,
\end{align}
from which the term $\mid \frac{1}{n h^2} \sum_{i=1}^n K^{\prime}\left(-\frac{\varepsilon_i}{h}\right) z_{i, j} z_{i, k}\mid $ can be further controlled as
$$
\begin{aligned}
& \mid \frac{1}{n h^2} \sum_{i=1}^n K^{\prime}\left(-\frac{\varepsilon_i}{h}\right) z_{i, j} z_{i, k}\mid  \\
\leq & \mid \frac{1}{n h^2} \sum_{i=1}^n K^{\prime}\left(-\frac{\varepsilon_i}{h}\right) z_{i, j} z_{i, k}-\mathbb{E}\left[\frac{1}{h^2} K^{\prime}\left(-\frac{\varepsilon_i}{h}\right) z_{i, j} z_{i, k}\right]\mid \notag\\
+&\mid \mathbb{E}\left[\frac{1}{h^2} K^{\prime}\left(-\frac{\varepsilon_i}{h}\right) z_{i, j} z_{i, k}\right]\mid  .
\end{aligned}
$$

Given $\boldsymbol{Z}_i$, the conditional mean $\mathbb{E}\left[\left.\frac{1}{h^2} K^{\prime}\left(-\frac{\varepsilon_i}{h}\right) \right\rvert\, \boldsymbol{Z}_i\right]$ satisfies
$$
\begin{aligned}
\mid \mathbb{E}\left[\left.\frac{1}{h^2} K^{\prime}\left(-\frac{\varepsilon_i}{h}\right) \right\rvert\, \boldsymbol{Z}_i\right]\mid  & =\frac{1}{h^2}\mid \int_{-\infty}^{\infty} K^{\prime}\left(-\frac{u}{h}\right) g_{\varepsilon_i \mid \boldsymbol{Z}_i}(u) \mathrm{d} u\mid  \\
& \leq \frac{1}{h} \int_{-\infty}^{\infty}\mid K^{\prime}(v)\mid  g_{\varepsilon_i \mid \boldsymbol{Z}_i}(-h v) \mathrm{d} v \leq \frac{2 \bar{g} C_K}{h}.
\end{aligned}
$$
where $C_K=\int_0^{\infty}\mid K^{\prime}(v)\mid  \mathrm{d} v<\infty$ is the total variation of $K(\cdot)$ on $[0, \infty)$. Therefore, we have
\begin{align} \label{eq6}
\mid \mathbb{E}\left[\frac{1}{h^2} K^{\prime}\left(-\frac{\varepsilon_i}{h}\right) z_{i, j} z_{i, k}\right]\mid  & \leq\mid \mathbb{E}\left[\mid \mathbb{E}\left[\left.\frac{1}{h^2} K^{\prime}\left(-\frac{\varepsilon_i}{h}\right) \right\rvert\, \boldsymbol{Z}_i\right]\mid  z_{i, j} z_{i, k}\right]\mid  \notag\\
& \leq \frac{2 \bar{g} C_K}{h}\left(\mathbb{E}\left(z_{i, j}\right)^2 \cdot \mathbb{E}\left(z_{i, k}\right)^2\right)^{\frac{1}{2}} \leq \frac{8 \bar{g} C_K \sigma^2}{h},
\end{align}
where the last inequality follows from $\mathbb{E}\left(z_{i, j}\right)^2 \leq 4 \sigma^2$ that has been verified in (A.1). For the centered term $\mid \frac{1}{n h^2} \sum_{i=1}^n K^{\prime}\left(-\frac{\varepsilon_i}{h}\right) z_{i, j} z_{i, k}-\mathbb{E}\left[\frac{1}{h^2} K^{\prime}\left(-\frac{\varepsilon_i}{h}\right) z_{i, j} z_{i, k}\right]\mid $, it can be bounded in a similar way to (A.3). Denote $\xi_i=\frac{1}{h^2} K^{\prime}\left(-\frac{\varepsilon_i}{h}\right)$, and then under Assumption 1 it can be verified that $\mid \xi_i\mid  \leq \frac{\kappa_u^{\prime}}{h^2}$ and
\begin{align}
    \mathbb{E}\left[\xi_i^2 \mid \boldsymbol{Z}_i\right]&=\frac{1}{h^4} \int_{-\infty}^{\infty} K^{\prime 2}\left(-\frac{u}{h}\right) g_{\varepsilon_i \mid \boldsymbol{Z}_i}(u) \mathrm{d} u\notag\\
    &=\frac{1}{h^3} \int_{-\infty}^{\infty} K^{\prime 2}(v) g_{\varepsilon_i \mid \boldsymbol{Z}_i}(-h v) \mathrm{d} v \notag\\
    &\leq \frac{2 \bar{g} \kappa_u^{\prime} C_K}{h^3} .
\end{align}

Analogously, applying Bernstein's inequality, we find that the following inequality holds with probability at least $1-\frac{1}{d}$,
$$
\begin{aligned}
& \max _{1 \leq j, k \leq d}\mid \frac{1}{n h^2} \sum_{i=1}^n K^{\prime}\left(-\frac{\varepsilon_i}{h}\right) z_{i, j} z_{i, k}-\mathbb{E}\left[\frac{1}{h^2} K^{\prime}\left(-\frac{\varepsilon_i}{h}\right) z_{i, j} z_{i, k}\right]\mid  \\
\leq & 8 \sigma^2\left(2 \sqrt{\bar{g} \kappa_u^{\prime} C_K \frac{\log d}{n h^3}}+\kappa_u^{\prime} \frac{\log d}{n h^2}\right) .
\end{aligned}
$$

Combining this bound with $r$, \eqref{equ1}, \eqref{eq5} and \eqref{eq6}, we have that with probability at least $1-\frac{3}{d}$,
\begin{align} \label{eq7}
 J_1 \notag
\le&s \cdot r \cdot 8 \sigma^2 \cdot\left\{\frac{\bar{g} k}{h}+2 \sqrt{\bar{g} K_u^{\prime} C_k \frac{\log d}{n h^3}}+K_u^{\prime} \frac{\log d}{n h^2}\right\} \notag\\
= & O_p\left(\operatorname{sr}\left\{\frac{1}{h}+\sqrt{\frac{\log d}{n h^3}}+\frac{\log d}{n h^2}\right\}\right).
\end{align}

Combing equation \eqref{eq8} and \eqref{eq7}, we have
$$
A_1=O_p\left(sr\left(\frac{1}{h}+2 \sqrt{ \frac{\log d}{n h^3}}+\frac{\log d}{n h^2}\right)+\frac{s^2 r^2\log (2nd^{2})}{h^3}\right) .
$$

Proof. \underline{Proof of Theorem 1}

Define $D(\boldsymbol{\theta})=\left\langle\nabla \widehat{Q}_h(\boldsymbol{\theta} )-\nabla \widehat{Q}_h(\boldsymbol{\theta}^*), \boldsymbol{\theta}-\boldsymbol{\theta}^*\right\rangle$. To obtain the convergence of $\widehat{\boldsymbol{\theta}}$, we use the convexity of the loss function to get the upper bound and the lower bound of $D(\widehat{\boldsymbol{\theta}})$, in terms of $\|\widehat{\boldsymbol{\theta}}-\boldsymbol{\theta}^{*}\|_{2}$.

{\textit{Step I: Upper bound of $D(\boldsymbol{\theta})$}.}

Recall the loss function $\widehat{Q}_h(\boldsymbol{\theta})+\lambda\|\widehat{\boldsymbol{\sigma}} \odot \boldsymbol{\theta}\|_1$. There exists a subgradient $\widehat{\boldsymbol{g}} \in \partial\|\widehat{\boldsymbol{\sigma}} \odot \widehat{\boldsymbol{\theta}}\|_1$ such that $\nabla_{\boldsymbol{\theta}} \widehat{Q}_h(\hat{\boldsymbol{\theta}})+\lambda \widehat{\boldsymbol{g}}=\mathbf{0}$. Let $\widehat{\boldsymbol{\delta}}=\widehat{\boldsymbol{\theta}}-\boldsymbol{\theta}^*$. By the definition of subgradient, we have
$$
\begin{aligned}
\left\langle\widehat{\boldsymbol{g}}, \boldsymbol{\theta}^*-\widehat{\boldsymbol{\theta}}\right\rangle & \leq\left\|\widehat{\boldsymbol{\sigma}} \odot\boldsymbol{\theta}^*\right\|_1-\|\widehat{\boldsymbol{\sigma}} \odot\widehat{\boldsymbol{\theta}}\|_1=\left\|\widehat{\boldsymbol{\sigma}} \odot\boldsymbol{\theta}_{\mathcal{S}}^*\right\|_1-\left\|\widehat{\boldsymbol{\sigma}} \odot(\widehat{\boldsymbol{\delta}}+\boldsymbol{\theta}^*)\right\|_1 \\
& =\left\|\widehat{\boldsymbol{\sigma}} \odot\boldsymbol{\theta}_{\mathcal{S}}^*\right\|_1-\left\|\widehat{\boldsymbol{\sigma}} \odot\widehat{\boldsymbol{\delta}}_{\mathcal{S}^c}\right\|_1-\left\|\widehat{\boldsymbol{\sigma}} \odot(\widehat{\boldsymbol{\delta}}_{\mathcal{S}}+\boldsymbol{\theta}_{\mathcal{S}}^*)\right\|_1\\ &\leq\left\|\widehat{\boldsymbol{\sigma}} \odot\widehat{\boldsymbol{\delta}}_{\mathcal{S}}\right\|_1-\left\|\widehat{\boldsymbol{\sigma}} \odot\widehat{\boldsymbol{\delta}}_{\mathcal{S}^c}\right\|_1\\
&\leq \|\widehat{\boldsymbol{\sigma}}\|_{\max} \left(\left\|\widehat{\boldsymbol{\delta}}_{\mathcal{S}}\right\|_1-\left\|\widehat{\boldsymbol{\delta}}_{\mathcal{S}^c}\right\|_1\right).
\end{aligned}
$$

\textbf{We first derive the upper bound.} We have
$$
\begin{aligned}
D(\widehat{\boldsymbol{\theta}})=& \left\langle\nabla \widehat{Q}_h(\widehat{\boldsymbol{\theta}} )-\nabla \widehat{Q}_h(\boldsymbol{\theta}^*), \widehat{\boldsymbol{\theta}}-\boldsymbol{\theta}^*\right\rangle \\
=& \left\langle\nabla \widehat{Q}_{h}(\widehat{\boldsymbol{\theta}})-\nabla\widehat{Q}_{h}(\boldsymbol{\theta}^*)+\nabla \widetilde{Q}_{h}(\boldsymbol{\theta}^*)-\nabla \widetilde{Q}_{h}(\boldsymbol{\theta}^*)+\nabla Q_{h}(\boldsymbol{\theta}^*)-\nabla Q_{h}(\boldsymbol{\theta}^*), \widehat{\boldsymbol{\theta}}-\boldsymbol{\theta}^*\right\rangle\\
=&\left\langle\lambda \hat{\boldsymbol{g}}, \boldsymbol{\theta}^*-\widehat{\boldsymbol{\theta}}\right\rangle+\left\langle\nabla \widehat{Q}_h(\boldsymbol{\theta}^*)-\nabla \widetilde{Q}_h(\boldsymbol{\theta}^*), \boldsymbol{\theta}^*-\widehat{\boldsymbol{\theta}}\right\rangle\\
&+\left\langle\nabla \widetilde{Q}_h(\boldsymbol{\theta}^*)-\nabla Q_h(\boldsymbol{\theta}^*), \boldsymbol{\theta}^*-\widehat{\boldsymbol{\theta}}\right\rangle+\left\langle\nabla Q_h(\boldsymbol{\theta}^*), \boldsymbol{\theta}^*-\widehat{\boldsymbol{\theta}}\right\rangle\\
\leq & \lambda(\left\|\widehat{\boldsymbol{\delta}}_{\mathcal{S}}\right\|_1-\left\|\widehat{\boldsymbol{\delta}}_{\mathcal{S}^c}\right\|_1)+\underbrace{\left\|\nabla \widehat{Q}_h(\boldsymbol{\theta}^*)-\nabla_{\theta} \widetilde{Q}_h(\boldsymbol{\theta}^*)\right\|_{\infty}}_{=I_{1}}\|\widehat{\boldsymbol{\delta}}\|_1\\
&+\underbrace{\left\|\nabla \widetilde{Q}_h(\boldsymbol{\theta}^*)-\nabla Q_h(\boldsymbol{\theta}^*)\right\|_{\infty}}_{=I_{2}}\|\widehat{\boldsymbol{\delta}}\|_1+\underbrace{\left\|\boldsymbol{\Sigma}^{-1 / 2} \nabla Q_h(\boldsymbol{\theta}^*)\right\|_2}_{= b_h^*}\|\widehat{\boldsymbol{\delta}}\|_{\boldsymbol{\Sigma}}.
\end{aligned}
$$
According to the Lemma \ref{lemma A.2}, we have $I_{1}=\left\|\nabla\widetilde{Q}_h(\boldsymbol{\theta}^*)-\nabla Q_h(\boldsymbol{\theta}^*)\right\|_{\infty}=O_p\left(\sqrt{\frac{\log 2d}{n}}+\frac{\log 2d}{n}\right)$. By Lemma \ref{lemma A.3}, we have
\begin{eqnarray}
I_{2}&=&\|\nabla_{\boldsymbol{\theta}} \widehat{Q}_h(\boldsymbol{\theta}^*)-\nabla_{\boldsymbol{\theta}} \widetilde{Q}_h(\boldsymbol{\theta}^*)\|_{\infty}\notag\\
&=&O_p\left(\sqrt{\frac{\log d}{n}}+\sqrt{\frac{1}{d}}+s\left(\sqrt{\frac{\log n}{d}}+\sqrt{\frac{(\log d)(\log n)}{n}}\right)\right).\notag
\end{eqnarray} 
Since we assume $\lambda \asymp s\sigma  \sqrt{\tau(1-\tau) \log d \log n/ n}$, thus there exists a $\lambda$ such that $\left\{\lambda\ge 4I_{1}\right\}$ and $\left\{\lambda\ge 4I_{2}\right\}$. According to Lemma \ref{lemma A.1}, $b_{h}^{*}\leq \frac{1}{2}L_{0}\zeta_{2}h^{2}$. Then, we have
\begin{align} 
    &\left\langle\nabla_{\boldsymbol{\theta}} \widehat{Q}_h(\widehat{\boldsymbol{\theta}})-\nabla_{\boldsymbol{\theta}} \widehat{Q}_h(\boldsymbol{\theta}^*), \widehat{\boldsymbol{\theta}}-\boldsymbol{\theta}^*\right\rangle \notag\\
 \leq &\lambda(\left\|\widehat{\boldsymbol{\delta}}_{\mathcal{S}}\right\|_1-\left\|\widehat{\boldsymbol{\delta}}_{\mathcal{S}^c}\right\|_1+\|\widehat{\boldsymbol{\delta}}\|_1 / 4+\|\widehat{\boldsymbol{\delta}}\|_1/4)+b_h^*\|\widehat{\boldsymbol{\delta}}\|_{\boldsymbol{\Sigma}} \notag\\ \label{equ3}
 \leq &\frac{\lambda}{2}(3\left\|\widehat{\boldsymbol{\delta}}_{\mathcal{S}}\right\|_1-\left\|\widehat{\boldsymbol{\delta}}_{\mathcal{S}^c}\right\|_1)+b_h^*\|\widehat{\boldsymbol{\delta}}\|_{\boldsymbol{\Sigma}} \\ 
 \leq &\frac{3}{2} s^{1 / 2} \lambda\|\widehat{\boldsymbol{\delta}}\|_2+b_h^*\|\widehat{\boldsymbol{\delta}}\|_{\boldsymbol{\Sigma}}\notag \\ \label{eq9} 
 \leq & \frac{1}{2}\left(3s^{1/2}\lambda+L_{0}\zeta_{2}h^{2}\eta_{1}^{1/4}\right)\|\widehat{\boldsymbol{\delta}}\|_{2},
\end{align}   
where $\eta_{1}$ is the largest eigenvalue of $\boldsymbol{\Sigma}$.

Since $\widehat{Q}_h(\boldsymbol{\theta})$ is convex, we have $\left\langle\nabla_{\boldsymbol{\theta}} \widehat{Q}_h(\widehat{\boldsymbol{\theta}})-\nabla_{\boldsymbol{\theta}} \widehat{Q}_h(\boldsymbol{\theta}^*), \widehat{\boldsymbol{\theta}}-\boldsymbol{\theta}^*\right\rangle\geq0$, and the term in \eqref{equ3} is also larger than 0. Therefore, $\left\|\widehat{\boldsymbol{\delta}}_{\mathcal{S}^c}\right\|_1\leq 3\left\|\widehat{\boldsymbol{\delta}}_{\mathcal{S}}\right\|_1+b_h^*\|\widehat{\boldsymbol{\delta}}\|_{\boldsymbol{\Sigma}}$, from which it follows that
\begin{align}\label{eq22}
    \|\widehat{\boldsymbol{\delta}}\|_1 \leq\|\widehat{\boldsymbol{\delta}}_S\|_1+\|\widehat{\boldsymbol{\delta}}_{S^{C}}\|_1 \leq 4 s^{1 / 2}\|\widehat{\boldsymbol{\delta}}\|_2+2 \lambda^{-1} b_h^*\|\widehat{\boldsymbol{\delta}}\|_{\boldsymbol{\Sigma}} \leq(4s^{1 / 2}+L_0 \zeta_2 \lambda^{-1} h^2\eta_1^{1 / 4})\|\widehat{\boldsymbol{\delta}}\|_{2}.
\end{align}

\textit{Step II: Lower bound of $D(\widehat{\boldsymbol{\theta}})$}.

Using the Taylor series expansion, we get
$$
\begin{aligned}
D(\hat{\boldsymbol{\theta}})=& \left\langle\nabla_{\boldsymbol{\theta}} \widehat{Q}_h(\hat{\boldsymbol{\theta}})-\nabla_{\boldsymbol{\theta}} \widehat{Q}_h(\boldsymbol{\theta}^*), \hat{\boldsymbol{\theta}}-\boldsymbol{\theta}^*\right\rangle \\
 =&\frac{1}{n} \sum_{i=1}^n(\bar{K}(\frac{\hat{\boldsymbol{Z}}^{\top}_{i} \hat{\boldsymbol{\theta}}-Y_{t}}{h})-\bar{K}(\frac{\hat{\boldsymbol{Z}}_{i}^{\top} \boldsymbol{\theta}^*-Y_{t}}{h})) \hat{\boldsymbol{Z}}_{i}^{\top}(\hat{\boldsymbol{\theta}}- \boldsymbol{\theta}^*) \\
 \geqslant& \frac{1}{n} \sum_{i=1}^n(\bar{K}(\frac{\hat{\boldsymbol{Z}}_{i}^{\top} \boldsymbol{\theta}^*-Y_{t}}{h})+\frac{1}{h}K(\frac{\hat{\boldsymbol{Z}}^{\top}_{i} \boldsymbol{\theta}^*-Y_{t}}{h}) (\hat{\boldsymbol{\theta}}-\boldsymbol{\theta}^*)^{\top}\hat{\boldsymbol{Z}}_{i}\notag\\
 &-\bar{K}(\frac{\hat{\boldsymbol{Z}}_{i}^{\top} \boldsymbol{\theta}^*-Y_{t}}{h})) \hat{\boldsymbol{Z}}_{i}^{\top}(\hat{\boldsymbol{\theta}}-\boldsymbol{\theta}^*)\\
= &\hat{\boldsymbol{\delta}}^{\top}(\frac{1}{n} \sum_{i=1}^n K_{h}(-\widehat{r}(\boldsymbol{\theta}^{*}))\hat{\boldsymbol{Z}}_{i} \hat{\boldsymbol{Z}}_{i}^{\top}-E\frac{1}{n} \sum_{i=1}^n K_{h}(-\varepsilon_{i})\boldsymbol{Z}_{i} \boldsymbol{Z}_{i}^{\top})\hat{\boldsymbol{\delta}}\notag\\
&+\hat{\boldsymbol{\delta}}^{\top} E\frac{1}{n} \sum_{i=1}^n K_{h}(-\varepsilon_{i})\boldsymbol{Z}_{i} \boldsymbol{Z}_{i}^{\top}\hat{\boldsymbol{\delta}}\\
=&A_{1}+A_{2}.
\end{aligned}
$$

It is easy to derive that
$$
\begin{aligned}
|A_{1}|&\leq \|\widehat{\boldsymbol{\delta}}\|_{1}\cdot \|\frac{1}{n} \sum_{i=1}^n K_{h}(-\widehat{r}(\boldsymbol{\theta}^{*}))\hat{\boldsymbol{Z}}_{i} \hat{\boldsymbol{Z}}_{i}^{\top}-E\frac{1}{n} \sum_{i=1}^n K_{h}(-\varepsilon_{i})\boldsymbol{Z}_{i} \boldsymbol{Z}_{i}^{\top}\|_{\infty}\\
&\leq s\cdot\|\widehat{\boldsymbol{\delta}}\|_{2}\cdot \|\frac{1}{n} \sum_{i=1}^n K_{h}(-\widehat{r}(\boldsymbol{\theta}^{*}))\hat{\boldsymbol{Z}}_{i} \hat{\boldsymbol{Z}}_{i}^{\top}-E\frac{1}{n} \sum_{i=1}^n K_{h}(-\varepsilon_{i})\boldsymbol{Z}_{i} \boldsymbol{Z}_{i}^{\top}\|_{\infty}.
\end{aligned}
$$
By Lemma \ref{lemma A.4}, we have 
$$
\begin{aligned}
&\frac{1}{n} \sum_{i=1}^n K_{h}(-\widehat{r}(\boldsymbol{\theta}^{*}))\hat{\boldsymbol{Z}}_{i} \hat{\boldsymbol{Z}}_{i}^{\top}-E\frac{1}{n} \sum_{i=1}^n K_{h}(-\varepsilon_{i})\boldsymbol{Z}_{i} \boldsymbol{Z}_{i}^{\top}\\
=&O_p(\left(sr\left(\frac{1}{h}+2 \sqrt{ \frac{\log d}{n h^3}}+\frac{\log d}{n h^2}\right)+\frac{s^2 r^2\log (2nd^{2})}{h^3}\right)).
\end{aligned}
$$
Furthermore, since $s\cdot \left(\frac{\log d \log n}{n}\right)^{1/5}=o(1)$, then $A_{1}=o(A_{2})$, and we have 
\begin{align}\label{equ10}
    D(\widehat{\boldsymbol{\theta}})\ge \frac{1}{2}\hat{\boldsymbol{\delta}}^{\top} E\frac{1}{n} \sum_{i=1}^n K_{h}(-\varepsilon_{i})\boldsymbol{Z}_{i} \boldsymbol{Z}_{i}^{\top}\hat{\boldsymbol{\delta}}\ge \frac{1}{2}\eta_{d}\|\widehat{\boldsymbol{\delta}}\|_{2}^{2}.
\end{align}

\textit{Step III:  Combining Lower and Upper Bounds.}

Combining equations \eqref{eq9} and \eqref{equ10}, we have
$$
    \frac{1}{2}\eta_{d}\|\hat{\boldsymbol{\delta}}\|_{2}^{2}\leq D(\hat{\boldsymbol{\theta}})\leq \frac{1}{2}\left(3s^{1/2}\lambda+L_{0}\zeta_{2}h^{2}\eta_{1}^{1/4}\right)\|\widehat{\boldsymbol{\delta}}\|_{2}.  
$$
Therefore, 
$$
\|\hat{\boldsymbol{\delta}}\|_{2}\leq \eta_{d}^{-1}(3s^{1 / 2}\lambda+L_0 \zeta_2  h^2\eta_1^{1 / 4}).
$$
Finally, by \eqref{eq22}, we have $\|\hat{\boldsymbol{\delta}}\|_{1}\leq \sqrt{s}\|\hat{\boldsymbol{\delta}}\|_{2}\leq s^{1/2}\eta_{d}^{-1}(3s^{1 / 2}\lambda+L_0 \zeta_2  h^2\eta_1^{1 / 4})$.

\section{Proof of Theorem 4.1}

We first provide some useful lemmas.

\begin{lemma}\label{lemma A.6}
    Under the null hypothesis $\boldsymbol{\beta}^{*}(\tau)=\mathbf{0}$, we have
$$
\begin{aligned}
\| S_{\tau}(\widehat{\boldsymbol{\gamma}};\widehat{\boldsymbol{u}},\widehat{\boldsymbol{f}})-  S_{\tau}\left(\boldsymbol{\gamma}^{*};\widehat{\boldsymbol{u}},\widehat{\boldsymbol{f}}\right) \|_{\infty} = O_p\left(s^{1 / 2} \lambda+h^2\right).
\end{aligned}
$$
\end{lemma}

{\textit{Proof of Lemma \ref{lemma A.6}}.}
It is easy to derive that
$$
\begin{aligned}
& S(\hat{\boldsymbol{\gamma}} ; \widehat{\boldsymbol{u}}, \widehat{\boldsymbol{f}})-S\left(\boldsymbol{\gamma}^* ; \widehat{\boldsymbol{u}}, \widehat{\boldsymbol{f}}\right) \\
& =\frac{1}{n} \sum_{i=1}^n\left[\bar{K}_h\left(\widehat{\boldsymbol{f}}_i^{\top}\widehat{\boldsymbol{\gamma}} -Y_{i}\right)-\tau\right] \widehat{\boldsymbol{u}}_{i}^*-\frac{1}{n} \sum_{i=1}^n\left[\bar{K}_h\left(\widehat{\boldsymbol{f}}_i^{\top} \boldsymbol{\gamma}^*-Y_{i}\right)-\tau\right] \widehat{\boldsymbol{u}}_{i}^* \\
& =\frac{1}{n} \sum_{i=1}^n\left[\bar{K}_h\left(\widehat{\boldsymbol{f}}_{i}^{\top} \widehat{\boldsymbol{\gamma}}-Y_{i}\right)-\bar{K}_h\left(\widehat{\boldsymbol{f}}_i^{\top} \boldsymbol{\gamma}^{*}-Y_{i}\right)\right] \widehat{\boldsymbol{u}}_{i}^* \\
& =\frac{1}{n} \sum_{i=1}^n\left[K_h\left(\widehat{\boldsymbol{f}}_{i}^{\top} \boldsymbol{\gamma}^{*}-Y_{i}\right) \widehat{\boldsymbol{f}}_i^{\top}\left(\widehat{\boldsymbol{\gamma}}-\boldsymbol{\gamma}^{*}\right)+K_h^{\prime}\left(\widehat{\boldsymbol{f}}_{i}^{\top} \boldsymbol{\gamma}^{*} -Y_{i}\right)\left(\widehat{\boldsymbol{f}}_i^{\top}\left(\widehat{\boldsymbol{\gamma}}-\boldsymbol{\gamma}^{*}\right)\right)^{2}\right] \widehat{\boldsymbol{u}}_{i}^*.
\end{aligned}
$$
By Theorem 3.1, we have $\left\|\widehat{\boldsymbol{\theta}}-\boldsymbol{\theta}^*\right\|_2=O_P\left(s^{1 / 2} \lambda+h^2\right)$, which implys $\left\|\widehat{\boldsymbol{\gamma}}-\boldsymbol{\gamma}^*\right\|_2=O_P\left(s^{1 / 2} \lambda+h^2\right)$. Remember that $\boldsymbol{f}$ in a weighted manner; that is, $\boldsymbol{F}^{\top} \mathbf{K}_\tau \boldsymbol{u}^{*}_{\cdot j}=\mathbf{0}$, for $j=1, \ldots, d$. We have
$$
S(\hat{\boldsymbol{\gamma}} ; \widehat{\boldsymbol{u}}, \widehat{\boldsymbol{f}})-S\left(\boldsymbol{\gamma}^* ; \widehat{\boldsymbol{u}}, \widehat{\boldsymbol{f}}\right)  = O_p\left(s^{1 / 2} \lambda+h^2\right).
$$

\begin{lemma}\label{lemma A.7}
    Under the null hypothesis $\boldsymbol{\beta}^{*}(\tau)=\mathbf{0}$, we have
$$
\begin{aligned}
\| S_{\tau}\left(\boldsymbol{\gamma}^{*};\widehat{\boldsymbol{u}},\widehat{\boldsymbol{f}}\right)-S_{\tau}\left(\boldsymbol{\gamma}^{*};\boldsymbol{u},\boldsymbol{f}\right) \|_{\infty}
=  O_p \left(s\left(\sqrt{\frac{\log n}{d}}+\sqrt{\frac{(\log d)(\log n)}{n}}\right)\right).
\end{aligned}
$$
\end{lemma}

{\textit{Proof of Lemma \ref{lemma A.7}}.} It is easy to see that
$$
\begin{aligned}
& S(\boldsymbol{\gamma}^* ; \widehat{\boldsymbol{u}}, \widehat{\boldsymbol{f}})-S\left(\boldsymbol{\gamma}^* ; \widehat{\boldsymbol{u}}, \widehat{\boldsymbol{f}}\right) \\
 =&\frac{1}{n} \sum_{i=1}^n\left[\bar{K}_h\left(\widehat{\boldsymbol{f}}_i^{\top}\boldsymbol{\gamma}^* -Y_{i}\right)-\tau\right] \widehat{\boldsymbol{u}}_{i}^*-\frac{1}{n} \sum_{i=1}^n\left[\bar{K}_h\left(f_i^{\top} \boldsymbol{\gamma}^*-Y_{i}\right)-\tau\right] \boldsymbol{u}_{i}^{*} \\
 =&\frac{1}{n} \sum_{i=1}^n\left[\bar{K}_h\left(\widehat{\boldsymbol{f}}_i^{\top}\boldsymbol{\gamma}^* -Y_{i}\right)-\tau\right] \widehat{\boldsymbol{u}}_{i}^*-\frac{1}{n} \sum_{i=1}^n\left[\bar{K}_h\left(f_i^{\top}\boldsymbol{\gamma}^* -Y_{i}\right)-\tau\right] \widehat{\boldsymbol{u}}_{i}^*\\
&+\frac{1}{n} \sum_{i=1}^n\left[\bar{K}_h\left(f_i^{\top}\boldsymbol{\gamma}^* -Y_{i}\right)-\tau\right] \widehat{\boldsymbol{u}}_{i}^*-\frac{1}{n} \sum_{i=1}^n\left[\bar{K}_h\left(f_i^{\top} \boldsymbol{\gamma}^*-Y_{i}\right)-\tau\right] \boldsymbol{u}_{i}^{*}\\
= & I_{1}+I_{2}.
\end{aligned}
$$

\textbf{We first consider $I_{2}$.} We have 
$$
\begin{aligned}
    I_2&=\frac{1}{n} \sum_{i=1}^n\left[\bar{K}_h\left(f_i^{\top} \boldsymbol{\gamma}^*-Y_{i}\right)-\tau\right]\left(\widehat{\boldsymbol{u}}_{i}^*-\widehat{\boldsymbol{u}}_{i}\right)\\
& \leq \mid \mathbb{E}\left[\frac{1}{n} \sum_{i=1}^n\left[\tau-\bar{K}_h(-\varepsilon)\right]\left(\widehat{\boldsymbol{u}}_{i}-\boldsymbol{u}_{i}^{*}\right)\right]\mid \\
&+\mid \left(1-\mathbb{E}\right)\left[\frac{1}{n} \sum_{i=1}^n\left[\tau-\bar{K}_h(-\varepsilon)\right]\left(\widehat{\boldsymbol{u}}_{i}-\boldsymbol{u}_{i}^{*}\right)\right]\mid .
\end{aligned}
$$
Applying Hoeffding's inequality for the term\\ $\mid \left(1-\mathbb{E}\right)\left[\frac{1}{n} \sum_{i=1}^n\left[\tau-\bar{K}_h(-\varepsilon)\right]\left(\widehat{\boldsymbol{u}}_{i}-\boldsymbol{u}_{i}^{*}\right)\right]\mid $ implies that for any $t \geq 0$,
$$
\begin{aligned} 
&\mathbb{P}\left( \mid \left(1-\mathbb{E}\right)\left[\frac{1}{n} \sum_{i=1}^n\left[\tau-\bar{K}_h(-\varepsilon)\right]\left(\widehat{\boldsymbol{u}}_{i}-\boldsymbol{u}_{i}^{*}\right)\right]\mid \geq t\mid \boldsymbol{Z}\right) \\
\leq & 2 \exp\left(-\frac{nt^{2}}{2\max\{\tau,1-\tau\}\max\mid \widehat{\boldsymbol{u}}_{i}-\boldsymbol{u}_{i}^{*}\mid }\right).
\end{aligned}
$$

Let $t=2\max\{\tau,1-\tau\}\max\mid \widehat{\boldsymbol{u}}_{i}-\boldsymbol{u}_{i}^{*}\mid \sqrt{\frac{\log ( d)}{ n}}$, then we obtain
$$
\mid \left(1-\mathbb{E}\right)\left[\frac{1}{n} \sum_{i=1}^n\left[\tau-\bar{K}_h(-\varepsilon)\right]\left(\widehat{\boldsymbol{u}}_{i}-\boldsymbol{u}_{i}^{*}\right)\right]\mid  \leq \sqrt{\frac{\log  d}{n}}.
$$
On the other hand, it can be verified that
$$
\begin{aligned}
\mathbb{E}\left[\bar{K}_h(-\varepsilon) \mid \boldsymbol{Z}\right] & =\int_{-\infty}^{\infty} \bar{K}_h\left(-\frac{s}{h}\right) \mathrm{d} G_{\varepsilon \mid \boldsymbol{Z}}(s)=-\frac{1}{h} \int_{-\infty}^{\infty} K\left(-\frac{s}{h}\right) G_{\varepsilon \mid \boldsymbol{Z}}(s) \mathrm{d} s \\
& =\int_{-\infty}^{\infty} K(v) G_{\varepsilon \mid \boldsymbol{Z}}(-h v) \mathrm{d} v\notag\\
&=\tau+\int_{-\infty}^{\infty} K(v) \int_0^{-h v}\left\{g_{\varepsilon \mid \boldsymbol{Z}}(t)-g_{\varepsilon \mid \boldsymbol{Z}}(0)\right\} \mathrm{d} t \mathrm{~d} v,
\end{aligned}
$$
from which it follows that $\mid \mathbb{E}\left[\bar{K}_h(-\varepsilon) \mid \boldsymbol{Z}\right]-\tau\mid  \leq \frac{1}{2} L_0 \kappa_2 h^2$. Consequently,
$$
\mid \mathbb{E}\left[\frac{1}{n} \sum_{i=1}^n\left[\tau-\bar{K}_h(-\varepsilon)\right]\left(\widehat{\boldsymbol{u}}_{i}-\boldsymbol{u}_{i}^{*}\right)\right]\mid  \leq \frac{1}{2} L_0 \kappa_2  h^2 .
$$

\textbf{We now consider $I_{1}$.} We have 
$$
\begin{aligned}
I_1 & =\frac{1}{n} \sum_{i=1}^n\left[\bar{K}_h\left(\widehat{\boldsymbol{f}}_i^{\top} \boldsymbol{\gamma}^*- Y_{i}\right)-\bar{K}_h\left(f_i^{\top} \boldsymbol{\gamma}^*-Y_{i}\right)\right] \widehat{\boldsymbol{u}}_{i}^*. 
\end{aligned}
$$
According to Lemma \ref{lemma A.3}, $I_{1} = O_p\left(s\left(\sqrt{\frac{\log n}{d}}+\sqrt{\frac{(\log d)(\log n)}{n}}\right)\right).$

Proof. \underline{Proof of Theorem 4.1}

By Lemma \ref{lemma A.6} and \ref{lemma A.7}, we have
$$
\begin{aligned}
    &\|S_{\tau}(\widehat{\boldsymbol{\gamma}};\widehat{\boldsymbol{u}},\widehat{\boldsymbol{f}})-S_{\tau}\left(\boldsymbol{\gamma}^{*};\boldsymbol{u},\boldsymbol{f}\right)\|_{\infty}\\
    =&\|S_{\tau}(\widehat{\boldsymbol{\gamma}};\widehat{\boldsymbol{u}},\widehat{\boldsymbol{f}})-S_{\tau}\left(\boldsymbol{\gamma}^{*};\widehat{\boldsymbol{u}},\widehat{\boldsymbol{f}}\right)+S_{\tau}\left(\boldsymbol{\gamma}^{*};\widehat{\boldsymbol{u}},\widehat{\boldsymbol{f}}\right)-S_{\tau}\left(\boldsymbol{\gamma}^{*};\boldsymbol{u},\boldsymbol{f}\right)\|_{\infty}\\
    =&O_p \left(s\left(\sqrt{\frac{\log n}{d}}+\sqrt{\frac{(\log d)(\log n)}{n}}\right)\right).
\end{aligned}
$$

\bibliography{sn-bibliography}% common bib file
